\documentclass[onecolumn]{revtex4}%
\usepackage{amssymb}
\usepackage{amsmath}
\usepackage{epsfig}
\usepackage{amsfonts}
\usepackage{graphicx}%
\begin{document}
\title{Study of chiral symmetry restoration in linear and nonlinear $O(N)$ models
using the auxiliary-field method}
\author{Elina Seel$^{\text{a}}$, Stefan Str\"{u}ber$^{\text{a}}$, Francesco
Giacosa$^{\text{a}}$, and Dirk H.\ Rischke$^{\text{a,b}}$}
\affiliation{$^{\text{a}}$Institute for Theoretical Physics, Goethe University,
Max-von-Laue-Str.\ 1, D--60438 Frankfurt am Main, Germany}
\affiliation{$^{\text{b}}$Frankfurt Institute for Advanced Studies, Goethe University,
Ruth-Moufang-Str.\ 1, D--60438 Frankfurt am Main, Germany }

\begin{abstract}
We consider the $O(N)$ linear $\sigma$ model and introduce an auxiliary field
to eliminate the scalar self-interaction. Using a suitable limiting process
this model can be continuously transformed into the nonlinear version of the
$O(N)$ model. We demonstrate that, up to two-loop order in the CJT formalism,
the effective potential of the model with auxiliary field is identical to the
one of the standard $O(N)$ linear $\sigma$ model, if the auxiliary field is
eliminated using the stationary values for the corresponding one- and
two-point functions. We numerically compute the chiral condensate and the
$\sigma-$ and $\pi-$meson masses at nonzero temperature in the one-loop
approximation of the CJT formalism. The order of the chiral phase transition
depends sensitively on the choice of the renormalization scheme. In the linear
version of the model and for explicitly broken chiral symmetry, it turns from
crossover to first order as the mass of the $\sigma$ particle increases. In
the nonlinear case, the order of the phase transition turns out to be of first
order. In the region where the parameter space of the model allows for
physical solutions, Goldstone's theorem is always fulfilled.

\end{abstract}
\maketitle

\section{Introduction}

Scalar models in $d+1$ space-time dimensions with orthogonal symmetry are
widely used in many areas of physics. Some applications of these $O(N)$ models
are quantum dots, high-temperature superconductivity, low-dimensional systems,
polymers, organic metals, biological molecular arrays, and chains. In this
paper, we focus on a physical system consisting of interacting pions and
$\sigma$ mesons at nonzero temperature $T$. For three spatial dimensions,
$d=3$, an analytical solution to this model does not exist. Thus, one has to
use many-body approximation schemes in order to compute quantities of
interest, such as the effective potential, the order parameter, and the masses
of the particles as a function of $T$. As an approximation scheme never gives
the exact solution, it is of interest to compare different schemes and assess
their physical relevance.

For $N=4$ the $O(N)$ symmetry group for the internal degrees of freedom is
locally isomorphic to the chiral $SU(2)_{R}\times SU(2)_{L}$ symmetry group of
quantum chromodynamics (QCD) with $N_{f} = 2$ massless quark flavors. The
phenomena of low-energy QCD are largely governed by chiral symmetry.

In the case of zero quark masses the QCD Lagrangian is invariant under
$U(N_{f})_{R}\times U(N_{f})_{L}$ transformations, $N_{f}$ being the number of
quark flavors. However, the true symmetry of QCD is only $U(N_{f})_{V}\times
SU(N_{f})_{A}$, because of the axial anomaly which explicitly breaks
$U(1)_{A}$ due nontrivial topological effects \cite{hooft}. For $N_{f}$
nonzero but degenerate quark masses, the $SU(N_{f})_{A}$ symmetry is
explicitly broken, such that QCD has only a $U(N_{f})_{V}$ flavor symmetry. In
reality, different quark flavors have different masses, reducing the symmetry
of QCD to $U(1)_{V}$, which corresponds to baryon number conservation. In the
vacuum, the axial $SU(N_{f})_{A}$ symmetry is also spontaneously broken by a
non-vanishing expectation value of the quark condensate $\left\langle q\bar
{q}\right\rangle \neq0$ \cite{witten}. According to Goldstone's theorem, this
leads to $N_{f}^{2}-1$ Goldstone bosons.

The chiral symmetry is restored at a temperature $T$ which for dimensional
reasons is expected to be of the order of $\Lambda_{QCD}\sim200$ MeV. This
scenario is indeed confirmed by lattice simulations, in which 
(for physical quark masses) a crossover
transition at $T_{c}\sim150$ MeV has been observed \cite{1}.

For vanishing quark masses, the high- and the low-temperature phases of QCD
have different symmetries, and therefore must be separated by a phase
transition. The order of this chiral phase transition is determined by the
global symmetry of the QCD Lagrangian; for $U(N_{f})_{V}\times U(N_{f})_{A}$,
the transition is of first order if $N_{f}\geq2$; for $U(N_{f})_{V}\times
SU(N_{f})_{A}$ the transition can be of second order if $N_{f}\leq2$
\cite{pisarski}. If the quark masses are nonzero, the second-order phase
transition becomes crossover.

The calculation of hadronic properties at nonzero temperature faces serious
technical difficulties. For a nonconvex effective potential standard
perturbation theory cannot be applied. Furthermore, nonzero
temperature introduces an
additional scale which invalidates the usual power counting in
terms of the coupling constant \cite{dolan}. A consistent calculation to a
given order in the coupling constant then may require a resummation of whole
classes of diagrams \cite{braaten}.

A convenient technique to perform such a resummation and thus arrive at a
particular many-body approximation scheme is the so-called two-particle
irreducible (2PI) or Cornwall-Jackiw-Tomboulis (CJT) formalism \cite{cornwall,
kadanoff}, which is a relativistic generalization of the $\Phi$-functional
formalism \cite{luttinger, baym}. The CJT formalism extends the concept of the
generating functional $\Gamma\left[  \phi\right]  $ for one-particle
irreducible (1PI) Green's functions to that for 2PI Green's functions
$\Gamma\left[  \phi,G\right]  $, where $\phi$ and $G$ are the one- and
two-point functions. The central quantity in this formalism is the sum of all
2PI vacuum diagrams, $\Gamma_{2}\left[  \phi,G\right]  $. Any many-body
approximation scheme can be derived as a particular truncation of $\Gamma
_{2}\left[  \phi,G\right]  $.

An advantage of the CJT formalism is that it avoids double counting and
fulfills detailed balance relations and thus is thermodynamically consistent.
Another advantage is that the Noether currents are conserved for an arbitrary
truncation of $\Gamma_{2}$, as long as the one- and two-point functions
transform as rank-1 and rank-2 tensors. A disadvantage is that Ward-Takahashi
identities for higher-order vertex functions are no longer fulfilled
\cite{hees}. As a consequence, Goldstone's theorem is violated
\cite{petropoulos,dirk}. A strategy to restore Goldstone's theorem is to
perform a so-called \textquotedblleft external\textquotedblright\ resummation
of random-phase approximation diagrams with internal lines given by the full
propagators of the approximation used in the CJT formalism \cite{hees}.

In the literature different many-body approximations have been applied to
examine the thermodynamical behavior of the $O(N)$ model in its linear and
nonlinear versions. In Ref.\ \cite{chiku} optimized perturbation theory was
used to compute the effective potential, spectral functions, and dilepton
emission rates. The CJT formalism has been applied to study the thermodynamics
of the $O(N)$ model in the so-called \textquotedblleft
double-bubble\textquotedblright\ approximation \cite{petropoulos, dirk,
bielich, roder, grinstein, pol, camelia, amelino, roh, nemoto, petro,
knollivanov}, in Ref.\ \cite{ruppert} sunset-type diagrams have been included.
The $1/N$ expansion has also been used several times to study various
properties of the $O(N)$ model at zero \cite{coleman, root} and nonzero
\cite{meyer, bochkarev, warringa, brauner} temperature.

In this paper, we derive the effective potential for the $O(N)$ linear
$\sigma$ model within the auxiliary-field method
\cite{Cooper:2005vw,Jakovac:2008zq,Fejos:2009dm}. 
The auxiliary field allows
us to obtain the nonlinear version of the $\sigma$ model by a well-defined
limiting process from the linear version. We demonstrate that, to two-loop
order, the effective potential is equivalent to the one of the standard $O(N)$
linear $\sigma$ model without auxiliary field, once the one- and two-point
functions involving the auxiliary field are replaced by their stationary
values. We then calculate the masses and the condensates of the $O(N)$ model
at nonzero $T$ in one-loop approximation. Although we restrict our treatment
to one-loop order, the condensate equation for the auxiliary field introduces
self-consistently computed loops in the equations for the masses. Therefore,
the one-loop approximation with auxiliary field is qualitatively similar to
the standard double-bubble (Hartree-Fock) approximation in the treatment
without auxiliary field. However, since the equations for the masses differ
quantitatively, they lead to different results for the order parameter and the
masses of the particles as a function of $T$.

The order of the chiral phase transition depends sensitively on the choice of
renormalization scheme. In the linear version of the model and for explicitly
broken chiral symmetry, it turns from crossover to first order as the mass of
the $\sigma$ particle increases. In the counter-term renormalization scheme,
this transition happens for smaller values of the $\sigma$ meson than in the
case where vacuum contributions to tadpole diagrams are simply neglected (the
so-called trivial regularization). In the nonlinear case the phase transition
is of first order. Besides, in the region where the parameter space of the
model allows for physical solutions of the mass equations, Goldstone's theorem
is always respected.

The manuscript is organized as follows: in Sec.\ II the linear and nonlinear
versions of the model are presented and it is shown 
how they can be related with the help
of an auxiliary field. In Sec.\ III the effective potential and the equations
for the condensate and masses are derived. We demonstrate the equivalence of
the auxiliary-field method to that of the standard approach (i.e., without
auxiliary field) when replacing the one- and two-point functions of the
auxiliary field by their stationary values. In Sec.\ IV the results are
presented for the linear and nonlinear versions of the model in the case of
non-vanishing and vanishing explicit symmetry breaking. Section V concludes
this paper with a summary of our results and an outlook for further studies.
An Appendix contains an alternative proof of the equivalence of the treatment
with and without auxiliary field, and details concerning the renormalization
of tadpole integrals.

We use units $h=c=k_{B}=1.$ The metric tensor is $g^{\mu\nu}=$
diag$(1,-1,-1,-1).$ Four-vectors are denoted by capital letters, $K^{\mu}=
(k_{0},\vec{k})$. We use the imaginary-time formalism to compute quantities at
nonzero temperature, i.e., the energy is $k_{0} = i \omega_{n}$, where
$\omega_{n}$ is the Matsubara frequency. For bosons, $\omega_{n} = 2 \pi n T$.
Energy-momentum integrals are denoted as
\begin{equation}
\int_{K}f(K)\equiv T\sum_{n=-\infty}^{\infty}\int\frac{d^{3}\vec{k}}{\left(
2\pi\right)  ^{3}}f(i\omega_{n},\vec{k})\text{ .}%
\end{equation}

\section{The $O(N)$ model}

The generating functional of the $\sigma$ model with $O(N)$ symmetry at
nonzero temperature $T$ is given by
\begin{equation}
Z_{L}(\varepsilon,h)=\mathcal{N}\int\mathcal{D}\alpha\,\mathcal{D}\Phi
\,\exp\left(  \int_{0}^{1/T}d\tau\int_{V}^{\ }d^{3}\vec{x}\,\mathcal{L}%
_{\sigma\text{-}\alpha}\right)  \;, \label{gf1}%
\end{equation}
with the Lagrangian
\begin{equation}%
\begin{array}
[c]{ccc}%
\mathcal{L}_{\sigma\text{-}\alpha}=\dfrac{1}{2}\left(  \partial_{\mu}%
\Phi\right)  ^{2}-U(\Phi,\alpha) & , & \text{ }U(\Phi,\alpha)=\dfrac{i}%
{2}\alpha(\Phi^{2}-\upsilon_{0}^{2})+\dfrac{N\varepsilon}{8}\alpha^{2}%
-h\sigma\text{ ,}%
\end{array}
\label{l1}%
\end{equation}
where $\Phi^{2}=\Phi^{t}\Phi$, $\Phi^{t}=\left(  \sigma,\pi_{1},\ldots,
\pi_{N-1}\right)  $ and $\alpha$ is an auxiliary field serving as a Lagrange
multiplier. One can obtain the generating functional of the $O(N)$ model in
its familiar form by integrating out the field $\alpha$:
\begin{equation}
Z_{L}(\varepsilon,h)=\tilde{\mathcal{N}}\int\mathcal{D}\Phi\,\exp\left(
\int_{0}^{1/T} d\tau\int_{V}d^{3}\vec{x}\,\mathcal{L}_{\sigma}\right)  \;,
\end{equation}
with the Lagrangian
\begin{equation}
\mathcal{L}_{\sigma}=\underset{\ }{\dfrac{1}{2}\left(  \partial_{\mu}
\Phi\right)  ^{2}-\frac{1}{2N\varepsilon}\left(  \Phi^{2}-\upsilon_{0}
^{2}\right)  ^{2}+h\sigma\text{ }.\text{ }} \label{lag}%
\end{equation}
As one can see, the potential of the model exhibits the typical tilted
Mexican-hat shape, with the parameter $1/\varepsilon$ being the coupling
constant, $h$ the parameter for explicit symmetry breaking, and $\upsilon_{0}$
the vacuum expectation value (v.e.v.) of $\Phi$. The $\pi_{i}$ fields can be
identified as the pseudo-Goldstone fluctuations.

Another way to see the equivalence to the standard form of the $O(N)$ model is
to use the equation of motion for the auxiliary field $\alpha$,
\begin{equation}%
\begin{array}
[c]{ccc}%
\dfrac{\delta\mathcal{L}_{\sigma\text{-}\alpha}}{\delta\alpha}-\partial_{\mu
}\dfrac{\delta\mathcal{L}_{\sigma\text{-}\alpha}}{\delta\partial_{\mu}\alpha
}=0\text{ }\  & \Longrightarrow & \text{ }i\alpha=\dfrac{2}{N\varepsilon}
(\Phi^{2}-\upsilon_{0}^{2})\text{ }.
\end{array}
\end{equation}
When plugging the latter into $\mathcal{L}_{\sigma\text{-}\alpha}$ one
recovers, as expected, the familiar Lagrangian $\mathcal{L}_{\sigma}$.

The advantage of the representation (\ref{gf1}) of the generating functional
of the \emph{linear\/} $\sigma$ model is that, by taking the limit
$\varepsilon\rightarrow0^{+}$, one naturally obtains the \emph{nonlinear\/}
version of the $\sigma$ model with the fields constrained by the condition
$\Phi^{2}=\upsilon_{0}^{2}$. In fact,
\begin{align}
Z_{NL}(h)  &  =\lim_{\varepsilon\rightarrow\text{ }0^{+}}Z_{L} (\varepsilon
,h)=\lim_{\varepsilon\rightarrow\text{ }0^{+}}\mathcal{N}\int\mathcal{D}
\alpha\,\mathcal{D}\Phi\,\exp\left[  \int_{0}^{1/T}d\tau\int_{V} d^{3}\vec
{x}\,\mathcal{L}_{\sigma\text{-}\alpha}\right] \nonumber\\
&  = \mathcal{N}^{\prime}\int\mathcal{D}\Phi\,\delta[\Phi^{2}-\upsilon_{0}%
^{2}]\exp\left\{  \int_{0}^{1/T}d\tau\int_{V}d^{3}\vec{x}\left[  \dfrac{1}%
{2}\left(  \partial_{\mu} \Phi\right)  ^{2}+h\sigma\right]  \right\}  \text{
}, \label{z2}%
\end{align}
because $\delta[\Phi^{2}-\upsilon_{0}^{2}]$ can be identified with
\begin{equation}
\delta[\Phi^{2}-\upsilon_{0}^{2}]\sim\lim_{\varepsilon\rightarrow\text{ }0^{+}
}\int\mathcal{D}\alpha\,\exp\left\{  -\int_{0}^{1/T}d\tau\int_{V} d^{3}\vec
{x}\,\left[  \dfrac{i}{2}\alpha(\Phi^{2}-\upsilon_{0}^{2})+\dfrac
{N\varepsilon}{8}\alpha^{2}\right]  \right\}  \;,\text{ } \label{delta}%
\end{equation}
which is the mathematically well-defined (i.e., convergent) form of the usual
representation of the functional $\delta-$function. Equation (\ref{delta})
ensures that the Mexican hat potential becomes infinitely steep and,
consequently, the mass of the radial degree of freedom infinite.

Note that in some previous studies of the $O(N)$ nonlinear $\sigma$ model
\cite{meyer, bochkarev}, the $\varepsilon$-dependence in Eq.\ (\ref{delta})
was not appropriately handled: there, the limit $\varepsilon\rightarrow0^{+}$
was exchanged with the functional $\mathcal{D}\alpha$ integration, effectively
setting $\varepsilon= 0$ in the exponent. This, however, is incorrect, since
the additional term $\sim\varepsilon\, \alpha^{2}$ is essential to establish
the link between the linear model and the nonlinear one. Without this term, an
integration over the auxiliary field does not give the correct potential of
the linear model. Thus, for a proper construction of the nonlinear limit of
the $O(N)$ model the $\varepsilon$-dependence must be included.

\section{The CJT effective potential}

\label{CJT}

In this work we study the thermodynamical behavior of the $O(N)$
linear $\sigma$
model, and in particular the temperature dependence of the masses of the modes
and of the condensate. To this end one has to apply methods that go beyond the
standard loop expansion which is not applicable when the effective potential
is not convex \cite{rivers}, as is the case here because of spontaneous chiral
symmetry breaking. A method that allows to compute quantities like the
effective potential, the masses, and the order parameter at nonzero
temperature is provided by the Cornwall-Jackiw-Tomboulis (CJT) formalism
\cite{cornwall}. In order to apply this method, we need to identify the
tree-level potential, the tree-level propagators, as well as the interaction
vertices from the underlying Lagrangian.

\subsection{Tree-level potential, tree-level propagators, and vertices}

\label{IIIA}

In our case, the fields occurring in the Lagrangian are $\sigma
,\mbox{\boldmath ${\bf \pi}$}\equiv(\pi_{1},\ldots,\pi_{N-1})^{t}$, as well as
the auxiliary field $\alpha$. In general, the fields $\sigma$ and $\alpha$
attain non-vanishing vacuum expectation values. In order to take this fact
into account, we perform a shift $\sigma\rightarrow\phi+\sigma$ and
$\alpha\rightarrow\alpha_{0}+\alpha$, respectively. This leaves the kinetic
terms in the Lagrangian (\ref{l1}) unchanged, while the potential becomes
\begin{equation}
U(\sigma+\phi,\mbox{\boldmath $\bf \pi$},\alpha+\alpha_{0})=\dfrac{i}%
{2}(\alpha_{0}+\alpha)(\sigma^{2}+\mbox{\boldmath
${\bf\pi}$}^{2}+2\sigma\phi+\phi^{2}-\upsilon_{0}^{2})+\dfrac{N\varepsilon}%
{8}(\alpha_{0}+\alpha)^{2}-h(\phi+\sigma)\text{ }, \label{u1}%
\end{equation}
In order to derive the Lagrangian from which we can read off the tree-level
potential, the tree-level propagators, and the interaction vertices, we use
the fact that linear terms in the fields vanish on account of the famous
tadpole cancellation which utilizes the definition of the vacuum expectation
values via the conditions
\begin{equation}
\frac{dU}{d\phi}\equiv0\;,\;\;\;\;\frac{dU}{d\alpha_{0}}\equiv0\;.
\end{equation}
The resulting expression for the Lagrangian reads
\begin{align}
\mathcal{L}_{\sigma\text{-}\alpha}  &  =\dfrac{1}{2}\left(  \partial_{\mu
}\sigma\right)  ^{2}+\dfrac{1}{2}\left(  \partial_{\mu}\mbox{\boldmath
${\bf \pi}$}\right)  ^{2}-\dfrac{i\alpha_{0}}{2}\,\sigma^{2}-\dfrac
{i\alpha_{0}}{2}\,\mbox{\boldmath
${\bf\pi}$}^{2}-\dfrac{1}{2}\,\frac{N\varepsilon}{4}\,\alpha^{2}-i\phi
\sigma\alpha\nonumber\\
&  -\dfrac{i}{2}\alpha(\sigma^{2}+\mbox{\boldmath ${\bf \pi}$}^{2}%
)-U(\phi,\alpha_{0})\;, \label{lag2a}%
\end{align}
where the tree-level potential is
\begin{equation}
U(\phi,\alpha_{0})=\dfrac{i}{2}\alpha_{0}(\phi^{2}-\upsilon_{0}^{2}%
)+\dfrac{N\varepsilon}{8}\alpha_{0}^{2}-h\phi\text{ }. \label{U}%
\end{equation}
There is a bilinear mixing term, $i\phi\sigma\alpha,$ which renders the mass
matrix non-diagonal in the fields $\sigma$ and $\alpha$.

We can think of two ways to treat this mixing term:

\begin{enumerate}
\item[(i)] we keep this term and allow for a non-diagonal propagator which
mutually transforms the fields $\sigma$ and $\alpha$ into each other.

\item[(ii)] we perform a shift,
\begin{equation}
\alpha\longrightarrow\alpha-4\dfrac{i\phi}{N\varepsilon}\,\sigma\text{ },
\label{shift}%
\end{equation}
which eliminates the bilinear term.
\end{enumerate}

In the following, we discuss the construction of the CJT effective potential
only for case (ii). The discussion of case (i) will be delegated to Appendix
\ref{casei} where we explicitly demonstrate that, to two-loop order, the
effective potential and the equations for the condensates and the masses are
the same as for case (ii) when quantities involving the auxiliary field are
replaced by their stationary values.

After the shift (\ref{shift}), the resulting expression for the Lagrangian
reads
\begin{align}
\bar{\mathcal{L}}_{\sigma\text{-}\alpha}  &  =\dfrac{1}{2}\left(
\partial_{\mu}\sigma\right)  ^{2}+\dfrac{1}{2}\left(  \partial_{\mu
}\mbox{\boldmath
${\bf \pi}$}\right)  ^{2}-\dfrac{1}{2}\left(  i\alpha_{0}+\dfrac{4\phi^{2}%
}{N\varepsilon}\right)  \sigma^{2}-\dfrac{1}{2}\left(  i\alpha_{0}\right)
\mbox{\boldmath
${\bf\pi}$}^{2}-\dfrac{1}{2}\,\frac{N\varepsilon}{4}\,\alpha^{2}\nonumber\\
&  -\dfrac{i}{2}\alpha(\sigma^{2}+\mbox{\boldmath ${\bf \pi}$}^{2}%
)-\dfrac{2\phi}{N\varepsilon}\,\sigma(\sigma^{2}%
+\mbox{\boldmath ${\bf \pi}$}^{2})-U(\phi,\alpha_{0})\text{ }. \label{lag2}%
\end{align}
From this expression, we can immediately read off the inverse tree-level
propagator matrix,
\begin{equation}
\bar{D}^{-1}(K;\phi,\alpha_{0})=\left(
\begin{array}
[c]{cccc}%
\bar{D}_{\alpha}^{-1} & 0 & 0 & \cdots\\
0 & \bar{D}_{\sigma}^{-1}(K;\phi,\alpha_{0}) & 0 & \cdots\\
0 & 0 & \bar{D}_{\pi}^{-1}(K;\alpha_{0}) & \\
\vdots & \vdots &  & \ddots
\end{array}
\right)  =\left(
\begin{array}
[c]{cccc}%
\displaystyle\frac{N\varepsilon}{4} & 0 & 0 & \cdots\\
0 & -K^{2}+i\alpha_{0}+\displaystyle\frac{4\phi^{2}}{N\varepsilon} & 0 &
\cdots\\
0 & 0 & -K^{2}+i\alpha_{0} & \\
\vdots & \vdots &  & \ddots
\end{array}
\right)  \;. \label{pro-alpha}%
\end{equation}
The shift (\ref{shift}) has the following consequences:

\begin{enumerate}
\item[(a)] the Jacobian associated with the transformation is unity, thus the
functional integration in Eq.\ (\ref{z2}) remains unaffected.

\item[(b)] it generates a term in the $\sigma$ mass, which diverges in the
limit $\varepsilon\rightarrow0^{+},$ see Eq.\ (\ref{pro-alpha}). This is
expected, since the $\sigma$ particle becomes infinitely heavy in the
nonlinear version of the $O(N)$ model.
\end{enumerate}

Finally, we identify the tree-level vertices from the Lagrangian
(\ref{lag2a}): there are two three-point vertices connecting the auxiliary
field $\alpha$ to either two $\sigma$ or two {\boldmath$\mathbf{\pi}$ }
fields, respectively. (These are the same vertices that also appear in case
(i), see Appendix \ref{casei}.) Furthermore, there is a three-point vertex
with three $\sigma$ fields, and one with one $\sigma$ and two
{\boldmath$\mathbf{\pi}$}--fields. These vertices are proportional to $\phi$.
(These vertices arise from the shift (\ref{shift}); they do not appear in case
(i), see Appendix \ref{casei}.)

\subsection{CJT effective potential}

The effective potential assumes the form
\begin{align}
\lefteqn{V_{\rm eff}(\phi,\alpha_{0},G)=U(\phi,\alpha_{0})+\frac{1}{2}\int
_{K}\left[  \ln G_{\alpha}^{-1}(K)+\ln G_{\sigma}^{-1}(K)+(N-1)\ln G_{\pi
}^{-1}(K)\right]  }\nonumber\\
&  +\frac{1}{2}\int_{K}\left[  \bar{D}_{\alpha}^{-1}G_{\alpha}(K)+\bar
{D}_{\sigma}^{-1}(K;\phi,\alpha_{0})G_{\sigma}(K)+(N-1)\bar{D}_{\pi}%
^{-1}(K;\alpha_{0})G_{\pi}(K)-(N+1)\right]  +V_{2}(\phi,G)\;, \label{Veff}%
\end{align}
The term $V_{2}(\phi,G)$ represents the sum of all two-particle irreducible
diagrams constructed from the three-point vertices in Eq.\ (\ref{lag2}). By
definition, these diagrams consist of at least two loops. The one- and
two-point functions are determined by the stationary conditions for the
effective potential
\begin{equation}
\dfrac{\delta V_{\mathrm{eff}}}{\delta\phi}=0\text{ },\ \ \dfrac{\delta
V_{\mathrm{eff}}}{\delta\alpha_{0}}=0\text{ },\text{ }\dfrac{\delta
V_{\mathrm{eff}}}{\delta G_{i}(K)}=0\text{ },\;\;\;\;i=\alpha\,,\;\sigma
\,,\;\pi_{1}\,,\;\ldots\,,\;\pi_{N-1}\;. \label{nlsm-st-cond}%
\end{equation}
This leads to the following equations for the condensates,%
\begin{align}
h  &  =i\alpha_{0}\phi+\dfrac{4\phi}{N\varepsilon}\int_{K}G_{\sigma}%
(K)+\frac{\delta V_{2}(\phi,G)}{\delta\phi}\text{ },\label{con1}\\
i\alpha_{0}  &  =\dfrac{2}{N\varepsilon}\left[  \phi^{2}-\upsilon_{0}^{2}%
+\int_{K}G_{\sigma}(K)+(N-1)\int_{K}G_{\pi}(K)\right]  \text{ . } \label{a1}%
\end{align}
For the two-point functions we obtain from Eq.\ (\ref{nlsm-st-cond})
the Dyson equations
\begin{equation}
G_{\alpha}^{-1}(K)=\bar{D}_{\alpha}^{-1}+\Pi_{\alpha}(K)\;,\;\;\;\;G_{\sigma
}^{-1}(K)=\bar{D}_{\sigma}^{-1}(K;\phi,\alpha_{0})+\Pi_{\sigma}%
(K)\;,\;\;\;\;G_{\pi}^{-1}(K)=\bar{D}_{\pi}^{-1}(K;\alpha_{0})+\Pi_{\pi}(K)\;,
\label{Gshifteda}%
\end{equation}
where the self-energies are%
\begin{equation}
\Pi_{i}(K)=2\,\frac{\delta V_{2}(\phi,G)}{\delta G_{i}(K)}\;,\;\;\;\;i=\alpha
\,,\;\sigma\,,\;\pi_{1}\,,\;\ldots\,,\;\pi_{N-1}\;. \label{1PI2}%
\end{equation}
In the following two subsections, we give the explicit expressions for the
condensate and mass equations in one- and two-loop approximation, respectively.

\subsection{One-loop approximation}

In one-loop approximation, $V_{2}\equiv0$. Equation (\ref{a1}) remains the
same while Eq.\ (\ref{con1}) simplifies to
\begin{equation}
h=i\alpha_{0}\phi+\dfrac{4\phi}{N\varepsilon}\int_{K}G_{\sigma}(K)=\frac
{2\phi}{N\varepsilon}\left[  \phi^{2}-\upsilon_{0}^{2}+3\int_{K}G_{\sigma
}(K)+(N-1)\int_{K}G_{\pi}(K)\right]  \;, \label{con}%
\end{equation}
where for the second equality we have used Eq.\ (\ref{a1}) to replace
$i\alpha_{0}$.\ For $V_{2}=0$, all self-energies are zero,
cf.\ Eq.\ (\ref{1PI2}), i.e., the full inverse two-point functions are
identical to the inverse tree-level propagators. From Eq.\ (\ref{Gshifteda})
one immediately sees that the two-point functions for $\sigma$ meson and pion
can be written in the form
\begin{equation}
G_{i}(K)=[\bar{D}_{i}^{-1}(K;\phi,\alpha)]^{-1}=(-K^{2}+M_{i}^{2}%
)^{-1}\;,\;\;\;\;i=\sigma\,,\;\pi\;, \label{Gshifted}%
\end{equation}
with the (squared) masses
\begin{align}
M_{\sigma}^{2}  &  =i\alpha_{0}+\frac{4\phi^{2}}{N\varepsilon}\equiv\frac
{2}{N\varepsilon}\left[  3\,\phi^{2}-\upsilon_{0}^{2}+\int_{K}G_{\sigma
}(K)+(N-1)\int_{K}G_{\pi}(K)\right]  \;,\label{Msigma}\\
M_{\pi}^{2}  &  =i\alpha_{0}\equiv\frac{2}{N\varepsilon}\left[  \phi
^{2}-\upsilon_{0}^{2}+\int_{K}G_{\sigma}(K)+(N-1)\int_{K}G_{\pi}(K)\right]
\;. \label{Mpi}%
\end{align}
For the second equalities we have used the condensate equation (\ref{a1}) to
replace $i\alpha_{0}$. Note that this introduces self-consistently computed
tadpole integrals into the equations for the masses.

Neglecting terms which are subleading in $1/N$ --- an approximation commonly
referred to as the large-$N$ (or Hartree) limit --- Eqs.\ (\ref{con}),
(\ref{Msigma}), and (\ref{Mpi}) reduce to
\begin{align}
h  &  =\phi\,M_{\pi}^{2} +\mathcal{O}(N^{-1})\text{ },\label{gapLN}\\
M_{\sigma}^{2}  &  =M_{\pi}^{2} +\dfrac{4\phi^{2}}{N\varepsilon}\text{
},\label{MsLN}\\
M_{\pi}^{2}  &  =\frac{2}{N\varepsilon}\left[  \phi^{2}-\upsilon_{0}^{2}%
+N\int_{k}G_{\pi}(k)\right]  +\mathcal{O}(N^{-1})\text{ }. \label{MpLN}%
\end{align}
Note that the condensate and the v.e.v.\ are $\sim\sqrt{N},$ i.e., $\phi
^{2}\sim\upsilon_{0}^{2}$ $\sim N.$

\subsection{Two-loop approximation}

\begin{figure}
[ptb]
\begin{center}
\includegraphics[
height=2.6411in,
width=2.8885in
]%
{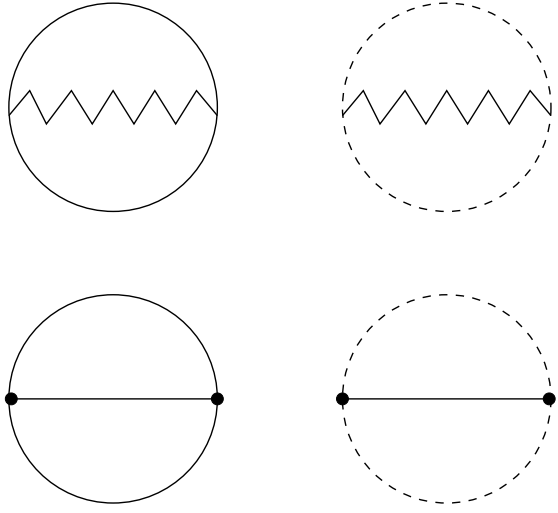}%
\caption{Two-particle irreducible diagrams constructed from the three-point
vertices in Eq.\ (\ref{lag2}). The full line represents the $\sigma$ field,
the dashed line represents the $\pi$ field and the zigzag line represents the
$\alpha$ field.}%
\label{V2(a)}%
\end{center}
\end{figure}

To two-loop order there are the four sunset-type diagrams shown in
Fig.\ \ref{V2(a)}, constructed from the three-point vertices between three
$\sigma$ fields, one $\sigma$ and two {\boldmath$\mathbf{\pi}$} fields, as
well as between one $\alpha$ field with either two $\sigma$ or two
{\boldmath$\mathbf{\pi}$} fields, respectively. There are no
double-bubble-type diagrams, due to the absence of four-point vertices. In
two-loop approximation,
\begin{align}
V_{2}(\phi,G)  &  =\frac{1}{4}\int_{K}\int_{P}G_{\alpha}(K+P)\left[
G_{\sigma}(K)G_{\sigma}(P)+(N-1)G_{\pi}(K)G_{\pi}(P)\right] \nonumber\\
&  -\left(  \dfrac{2\phi}{N\varepsilon}\right)  ^{2}\int_{K}\int_{P}G_{\sigma
}(K+P)\left[  3\,G_{\sigma}(K)G_{\sigma}(P)+(N-1)G_{\pi}(K)G_{\pi}(P)\right]
\text{ .} \label{V2}%
\end{align}
The overall sign follows from the fact that the effective potential has the
same sign as the free energy. The combinatorial factors in front of the
individual terms follow as usual from counting the possibilities of connecting
lines between the vertices, with an overall factor of $1/2$ because there are
two vertices.

The condensate equation (\ref{a1}) for the auxiliary field again remains
unchanged while Eq.\ (\ref{con1}) becomes
\begin{align}
h  &  = i\alpha_{0}\phi+\dfrac{4\phi}{N\varepsilon}\int_{K}G_{\sigma}(K) - 2
\phi\left(  \dfrac{2}{N\varepsilon}\right)  ^{2}\int_{K}\int_{P}G_{\sigma
}(K+P)\left[  3\,G_{\sigma}(K)G_{\sigma}(P)+(N-1)G_{\pi}(K)G_{\pi}(P)\right]
\nonumber\\
&  = \frac{2\phi}{N\varepsilon}\left\{  \phi^{2}-\upsilon_{0}^{2}+3\int
_{K}G_{\sigma}(K)+(N-1)\int_{K}G_{\pi}(K) \right. \nonumber\\
&  \hspace*{1cm} - \left.  \dfrac{4}{N\varepsilon}\int_{K}\int_{P}G_{\sigma
}(K+P)\left[  3\,G_{\sigma}(K)G_{\sigma}(P)+(N-1)G_{\pi}(K)G_{\pi}(P)\right]
\right\}  \text{ ,} \label{con12loop}%
\end{align}
where in the second equality we have used Eq.\ (\ref{a1}) to replace $i
\alpha_{0}$. This is identical with the condensate equation in the two-loop
approximation for the usual $O(N)$ linear $\sigma$ model without auxiliary
field, see Sec.\ \ref{recover}.

From Eq.\ (\ref{1PI2}) we derive the self-energies as
\begin{align}
\Pi_{\alpha} &  =\frac{1}{2}\int_{P}\left[  G_{\sigma}(P)G_{\sigma
}(K-P)+(N-1)\,G_{\pi}(P)G_{\pi}(K-P)\right]  \;,\\
\Pi_{\sigma}(K) &  =\int_{P}G_{\sigma}(P)G_{\alpha}(K-P)-2\left(  \dfrac
{2\phi}{N\varepsilon}\right)  ^{2}\int_{P}\left[  9\,G_{\sigma}(P)G_{\sigma
}(K-P)+(N-1)\,G_{\pi}(P)G_{\pi}(K-P)\right]  \;,\\
\Pi_{\pi}(K) &  =\int_{P}G_{\pi}(P)G_{\alpha}(K-P)-4\left(  \dfrac{2\phi
}{N\varepsilon}\right)  ^{2}\int_{P}G_{\sigma}(P)G_{\pi}(K-P)\;.
\end{align}
Then, the Dyson equations (\ref{Gshifteda}) for the full two-point functions
read
\begin{align}
G_{\alpha}^{-1}(K) &  =\bar{D}_{\alpha}^{-1}+\Pi_{\alpha}(K)=\frac
{N\varepsilon}{4}+\frac{1}{2}\int_{P}\left[  G_{\sigma}(P)G_{\sigma
}(K-P)+(N-1)\,G_{\pi}(P)G_{\pi}(K-P)\right]  \;,\label{Galpha}\\
G_{\sigma}^{-1}(K) &  =\bar{D}_{\sigma}^{-1}(K;\phi,\alpha_{0})+\Pi_{\sigma
}(K)=-K^{2}+i\alpha_{0}+\frac{4\phi^{2}}{N\varepsilon}+\Pi_{\sigma
}(K)\nonumber\\
&  =-K^{2}+\frac{2}{N\varepsilon}\left[  3\phi^{2}-\upsilon_{0}^{2}+\int
_{K}G_{\sigma}(K)+(N-1)\int_{K}G_{\pi}(K)\right]  +\int_{P}G_{\sigma
}(P)G_{\alpha}(K-P)\nonumber\\
&  -2\left(  \dfrac{2\phi}{N\varepsilon}\right)  ^{2}\int_{P}\left[
9\,G_{\sigma}(P)G_{\sigma}(K-P)+(N-1)\,G_{\pi}(P)G_{\pi}(K-P)\right]
\;,\label{Gsigma}\\
G_{\pi}^{-1}(K) &  =\bar{D}_{\pi}^{-1}(K;\alpha_{0})+\Pi_{\pi}(K)=-K^{2}%
+i\alpha_{0}+\Pi_{\pi}(K)\nonumber\\
&  =-K^{2}+\frac{2}{N\varepsilon}\left[  \phi^{2}-\upsilon_{0}^{2}+\int
_{K}G_{\sigma}(K)+(N-1)\int_{K}G_{\pi}(K)\right]  +\int_{P}G_{\pi}%
(P)G_{\alpha}(K-P)\nonumber\\
&  -4\left(  \dfrac{2\phi}{N\varepsilon}\right)  ^{2}\int_{P}G_{\sigma
}(P)G_{\pi}(K-P)\;.\label{Gpi}%
\end{align}
Here, we have also made use of Eq.\ (\ref{a1}) for the auxiliary field. %

\subsection{Recovering the standard two-loop approximation}

\label{recover}

In this subsection, we demonstrate that, to two-loop order, the results are
the same as for a direct application of the CJT formalism to the original
Lagrangian (\ref{lag}) of the $O(N)$ linear $\sigma$ model (a case that we
term ``standard two-loop approximation''), if we eliminate the $\alpha$ field
using the stationary values for the condensate $\alpha_{0}$ and the full
propagator $G_{\alpha}$. The effective potential for the original $O(N)$
linear $\sigma$ model reads
\begin{equation}
V_{\mathrm{eff}}^{\mathrm{l\sigma m}}(\phi,G)=\dfrac{1}{2N\varepsilon}%
(\phi^{2}-\upsilon_{0}^{2})^{2}-h\phi+\dfrac{1}{2}\sum_{i=\sigma
,\mbox{\scriptsize \boldmath ${\bf \pi}$}}\int_{K}[\ln G_{i}^{-1}%
(K)+D_{i}^{-1}(K;\phi)G_{i}(K)-1] +V_{2}^{\mathrm{l \sigma m}}(\phi,G)\;,
\label{VCJT}%
\end{equation}
where the inverse tree-level propagators are
\begin{equation}
\label{treelevelprop}D_{\sigma}^{-1}(K;\phi) = - K^{2} + \frac{2}%
{N\varepsilon}(3 \phi^{2} - \upsilon_{0}^{2})\;,\;\;\;\; D_{\pi}^{-1}(K;\phi)
= - K^{2} + \frac{2}{N\varepsilon} (\phi^{2} - \upsilon_{0}^{2})\;,
\end{equation}
and, to two-loop order,
\begin{align}
V_{2}^{\mathrm{l \sigma m}}(\phi, G)  &  = \dfrac{3}{2N\varepsilon}\left[
\int_{K}G_{\sigma}(K)\right]  ^{2}+(N+1)\dfrac{N-1}{2N\varepsilon}\left[
\int_{K}G_{\pi}(K)\right]  ^{2}+\dfrac{N-1}{N\varepsilon}\int_{K}G_{\pi
}(K)\int_{P}G_{\sigma}(P)\nonumber\\
&  - \left(  \dfrac{2\phi}{N\varepsilon}\right)  ^{2}\int_{K}\int_{P}%
G_{\sigma}(K+P)\left[  3\,G_{\sigma}(K)G_{\sigma}(P)+(N-1)G_{\pi}(K)G_{\pi
}(P)\right]  \;. \label{v2cjt}%
\end{align}
The first line is the contribution from double-bubble diagrams arising from
the four-point vertices with four $\sigma$ fields or two $\sigma$ and two
{\boldmath $\mathbf{\pi}$} fields in the Lagrangian (\ref{lag}). The second
line corresponds to the sunset diagrams shown in the second row of
Fig.\ \ref{V2(a)}. These are the same in the linear $\sigma$ model with or
without auxiliary field. Note that the sunset contribution differs in sign
from the double-bubble contribution [this sign was missed in
Ref.\ \cite{ruppert}]. The equation arising from the stationarity condition
(\ref{nlsm-st-cond}) for $V_{\mathrm{eff}}^{\mathrm{l \sigma m}}$ reads
\begin{align}
h  &  = \frac{2\phi}{N\varepsilon}\left\{  \phi^{2}-\upsilon_{0}^{2}+3\int
_{K}G_{\sigma}(K)+(N-1)\int_{K}G_{\pi}(K) \right. \nonumber\\
&  \hspace*{1cm} - \left.  \dfrac{4}{N\varepsilon}\int_{K}\int_{P}G_{\sigma
}(K+P)\left[  3\,G_{\sigma}(K)G_{\sigma}(P)+(N-1)G_{\pi}(K)G_{\pi}(P)\right]
\right\}  \text{ .}%
\end{align}
This is identical with Eq.\ (\ref{con12loop}), i.e., with the equation
obtained via the auxiliary-field formalism, once the auxiliary field is
eliminated with the help of Eq.\ (\ref{a1}).

The self-energies for $\sigma$ meson and pion read
\begin{align}
\Pi_{\sigma}^{\mathrm{l \sigma m}}(K)  &  = \frac{2}{N\varepsilon} \left[
3\int_{K} G_{\sigma}(K) + (N-1) \int_{K} G_{\pi}(K)\right] \nonumber\\
&  - 2\,\left(  \frac{2 \phi}{N\varepsilon} \right)  ^{2} \int_{P} \left[
9\,G_{\sigma}(P)G_{\sigma}(K-P) +(N-1)\,G_{\pi}(P)G_{\pi}(K-P) \right]  \;,\\
\Pi_{\pi}^{\mathrm{l \sigma m}}(K)  &  = \frac{2}{N\varepsilon} \left[
\int_{K} G_{\sigma}(K) + (N+1) \int_{K} G_{\pi}(K)\right]  - 4\,\left(
\frac{2 \phi}{N\varepsilon} \right)  ^{2} \int_{P} G_{\pi}(P)G_{\pi}(K-P)\;.
\end{align}
Therefore, the Dyson equations for the full two-point functions read
\begin{align}
G_{\sigma}^{-1}(K)  &  = D_{\sigma}^{-1}(K;\phi) + \Pi_{\sigma}^{\mathrm{l
\sigma m}}(K)\nonumber\\
&  = - K^{2} + \frac{2}{N\varepsilon}(3 \phi^{2} - \upsilon_{0}^{2}) +
\frac{2}{N\varepsilon} \left[  3\int_{K} G_{\sigma}(K) + (N-1) \int_{K}
G_{\pi}(K)\right] \nonumber\\
&  - 2\,\left(  \frac{2 \phi}{N\varepsilon} \right)  ^{2} \int_{P} \left[
9\,G_{\sigma}(P)G_{\sigma}(K-P) +(N-1)\,G_{\pi}(P)G_{\pi}(K-P) \right]
\;,\label{Gsigmalsm}\\
G_{\pi}^{-1}(K)  &  = D_{\pi}^{-1}(K;\phi) + \Pi_{\pi}^{\mathrm{l \sigma m}%
}(K)\nonumber\\
&  = - K^{2} + \frac{2}{N\varepsilon}(\phi^{2} - \upsilon_{0}^{2}) +\frac
{2}{N\varepsilon} \left[  \int_{K} G_{\sigma}(K) + (N+1) \int_{K} G_{\pi
}(K)\right]  - 4\,\left(  \frac{2 \phi}{N\varepsilon} \right)  ^{2} \int_{P}
G_{\pi}(P)G_{\sigma}(K-P)\;. \label{Gpilsm}%
\end{align}
These equations are identical with the Dyson equations (\ref{Gsigma}) and
(\ref{Gpi}), if we replace the propagator $G_{\alpha}$ of the auxiliary field
in those equations using the Dyson equation (\ref{Galpha}). In order to see
this, we formally write
\begin{equation}
\label{Galphaexp}G_{\alpha}(K) = \left[  G_{\alpha}^{-1}(K)\right]  ^{-1} =
\left[  \bar{D}_{\alpha}^{-1}+ \Pi_{\alpha}(K)\right]  ^{-1} = \bar{D}%
_{\alpha}\sum_{n=0}^{\infty}\left[  - \bar{D}_{\alpha}\Pi_{\alpha}(K)\right]
^{n}\;.
\end{equation}
If we insert this into the respective terms in Eqs.\ (\ref{Gsigma}) and
(\ref{Gpi}), we observe that the terms for $n \geq1$ generate contributions
which are at least of second order in loops (because $\Pi_{\alpha}(K)$ is
already a one-loop term). However, to two-loop order in the effective
potential, it is sufficient to consider the 1PI self-energies to one-loop
order only. Therefore, we may neglect all contributions in
Eq.\ (\ref{Galphaexp}) except for the $n=0$ (tree-level) term. Then, we may
replace
\begin{equation}
\int_{P} G_{i}(P) G_{\alpha}(K-P) \longrightarrow\int_{P} G_{i}(P) \bar
{D}_{\alpha}= \frac{4}{N\varepsilon} \int_{P} G_{i}(P)\;,\;\;\;\; i = \sigma,
\pi\;,
\end{equation}
in Eqs.\ (\ref{Gsigma}) and (\ref{Gpi}), i.e., they become simple tadpole
contributions to the self-energies. Combining these with the other tadpole
contributions, we observe that, indeed, Eqs.\ (\ref{Gsigma}) and (\ref{Gpi})
become identical with Eqs.\ (\ref{Gsigmalsm}) and (\ref{Gpilsm}).

Finally, we also show that the effective potential (\ref{V2}) in two-loop
approximation for $V_{2}(\phi,G)$, Eq.\ (\ref{V2}), becomes identical with the
effective potential for the standard linear $\sigma$ model, Eq.\ (\ref{v2cjt}%
), if we replace the expectation value and the full two-point function for the
auxiliary field by their stationary values. To this end, it is advantageous to
consider the tree-level, the one-loop, and the two-loop contributions in
Eq.\ (\ref{Veff}) separately. The tree-level potential at the stationary value
for $\alpha_{0}$ reads
\begin{align}
\lefteqn{U(\phi,\alpha_{0})=\frac{1}{2}\,(\phi^{2}-\upsilon_{0}^{2})\,\frac
{2}{N\varepsilon}\left[  \phi^{2}-\upsilon_{0}^{2}+\int_{K}G_{\sigma
}(K)+(N-1)\int_{K}G_{\pi}(K)\right]  }\nonumber\\
\lefteqn{\hspace*{1.7cm}-\frac{N\varepsilon}{8}\,\left(  \frac{2}%
{N\varepsilon}\right)  ^{2}\left[  \phi^{2}-\upsilon_{0}^{2}+\int_{K}%
G_{\sigma}(K)+(N-1)\int_{K}G_{\pi}(K)\right]  ^{2}-h\phi}\nonumber\\
&  =\frac{1}{2N\varepsilon}\left\{  \phi^{2}-\upsilon_{0}^{2}-\left[  \int
_{K}G_{\sigma}(K)\right]  ^{2}-2(N-1)\int_{K}G_{\sigma}(K)\int_{P}G_{\pi
}(P)-(N-1)^{2}\left[  \int_{K}G_{\pi}(K)\right]  ^{2}\right\}  -h\phi\,.
\label{treelevel}%
\end{align}
For the one-loop contribution, we expand the logarithm of the inverse
two-point function for the auxiliary field using the Dyson equation
(\ref{Galpha}) and employ the expansion (\ref{Galphaexp}),
\begin{align}
\lefteqn{\ln G_{\alpha}^{-1}(K)+\bar{D}_{\alpha}^{-1}G_{\alpha}(K)-1=\ln
\bar{D}_{\alpha}^{-1}+\ln\left[  1+\bar{D}_{\alpha}\Pi_{\alpha}(K)\right]
+\bar{D}_{\alpha}^{-1}\left[  \bar{D}_{\alpha}^{-1}+\Pi_{\alpha}(K)\right]
^{-1}-1}\nonumber\\
&  =\ln\frac{N\varepsilon}{4}+\bar{D}_{\alpha}\Pi_{\alpha}(K)-\sum
_{n=2}^{\infty}\frac{1}{n}\left[  -\bar{D}_{\alpha}\Pi_{\alpha}(K)\right]
^{n}+1-\bar{D}_{\alpha}\Pi_{\alpha}(K)+\sum_{n=2}^{\infty}\left[  -\bar
{D}_{\alpha}\Pi_{\alpha}(K)\right]  ^{n}-1\nonumber\\
&  =\ln\frac{N\varepsilon}{4}+\sum_{n=2}^{\infty}\left[  -\bar{D}_{\alpha}%
\Pi_{\alpha}(K)\right]  ^{n}\left(  1-\frac{1}{n}\right)  \;. \label{1loopexp}%
\end{align}
We observe that the terms linear in $\Pi_{\alpha}(K)$ as well as the unit
terms cancel. In the final result, the first term is a (negligible) constant.
The remaining series starts with a term with two powers of $\Pi_{\alpha}(K)$.
Since $\Pi_{\alpha}(K)$ is (at least) of one-loop order, when integrating over
$K$, this term is (at least) of three-loop order in the effective potential.
(In fact, since $\bar{D}_{\alpha}=4/(N\varepsilon)=const.$, one readily
convinces oneself that the $n=2$ term in the series corresponds to the
well-known basketball diagram.) To two-loop order in the effective potential,
we may therefore neglect the series in Eq.\ (\ref{1loopexp}).

Using Eq.\ (\ref{a1}), the remaining one-loop terms in the effective potential
(\ref{Veff}) read
\begin{align}
\lefteqn{\frac{1}{2}\int_{K}\left[  \ln G_{\sigma}^{-1}(K)+(N-1)\ln G_{\pi
}^{-1}(K)+\bar{D}_{\sigma}^{-1}(K;\phi,\alpha_{0})G_{\sigma}(K)+(N-1)\bar
{D}_{\pi}^{-1}(K;\alpha_{0})G_{\pi}(K)-N\right]  }\nonumber\\
&  =\frac{1}{2}\int_{K}\left\{  \ln G_{\sigma}^{-1}(K)+(N-1)\ln G_{\pi}%
^{-1}(K)\right. \nonumber\\
&  \hspace*{1cm}+\left.  \left[  -K^{2}+\frac{2}{N\varepsilon}\left(
3\phi^{2}-\upsilon_{0}^{2}\right)  \right]  G_{\sigma}(K)+(N-1)\left[
-K^{2}+\frac{2}{N\varepsilon}\left(  \phi^{2}-\upsilon_{0}^{2}\right)
\right]  G_{\pi}(K)-N\right\} \nonumber\\
&  +\frac{1}{N\varepsilon}\left[  \int_{K}G_{\sigma}(K)+(N-1)\int_{K}G_{\pi
}(K)\right]  ^{2}\;, \label{2loop}%
\end{align}
where the last term arises from the tadpole contributions to Eq.\ (\ref{a1}).
Multiplying them with full two-point functions $G_{\sigma}(K)$, $G_{\pi}(K)$,
and integrating over $K$, they lead to the double-bubble-type terms shown in
the last line. Note that the coefficients of the full two-point functions in
the second line are just the inverse tree-level propagators in the standard
linear $\sigma$ model, cf.\ Eq.\ (\ref{treelevelprop}).

Finally, we consider the two-loop contribution (\ref{V2}). To two-loop order,
it is justified to replace $G_{\alpha}(K+P) \rightarrow\bar{D}_{\alpha}%
\equiv4/(N\varepsilon)$, and we obtain
\begin{align}
V_{2}(\phi,G)  &  \simeq\frac{1}{N \varepsilon} \left\{  \left[  \int_{K}
G_{\sigma}(K) \right]  ^{2} +(N-1) \left[  \int_{K} G_{\pi}(K)\right]  ^{2}
\right\} \nonumber\\
&  - \left(  \dfrac{2\phi}{N\varepsilon}\right)  ^{2}\int_{K}\int_{P}%
G_{\sigma}(K+P)\left[  3\,G_{\sigma}(K)G_{\sigma}(P)+(N-1)G_{\pi}(K)G_{\pi
}(P)\right]  \text{ .}%
\end{align}
Adding Eqs.\ (\ref{treelevel}), (\ref{1loopexp}), and (\ref{2loop}), we indeed
obtain the effective potential (\ref{v2cjt}) of the standard linear $\sigma$ model.

\section{Results}

%

\begin{figure}
[ptb]
\begin{center}
\includegraphics[
height=2.399in,
width=5.5512in
]%
{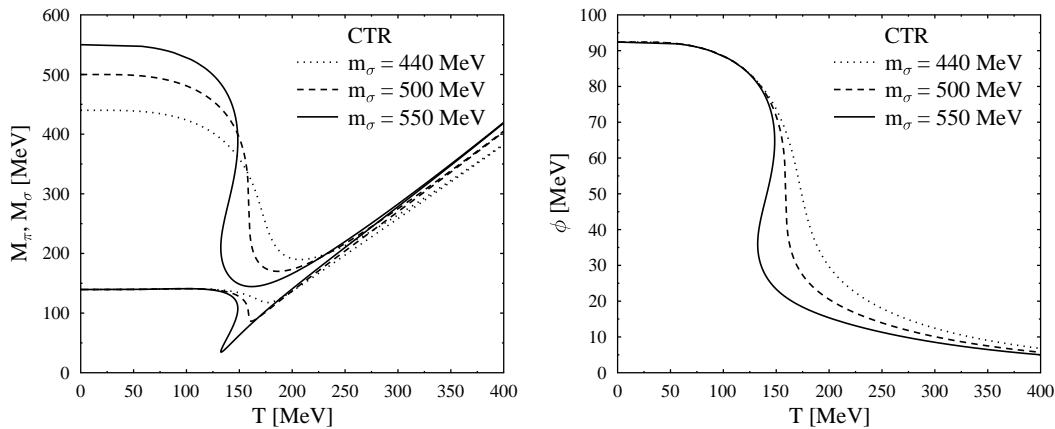}%
\caption{The pion mass, the sigma mass, and the condensate as a function of
$T$ in the $O(4)$ linear $\sigma$ 
model in case of explicitly broken symmetry using the
CTR\ scheme for different values of $m_{\sigma}$.}%
\label{lin-ct}%
\end{center}
\end{figure}

In this section, we show numerical solutions for the one-loop approximation,
Eqs.\ (\ref{con}), (\ref{Msigma}), and (\ref{Mpi}), for $N=4$, corresponding
to a system of three pions and their chiral partner, the $\sigma$ particle. We
compare this to results for the one-loop approximation in the large--$N$
limit, Eqs.\ (\ref{gapLN}) -- (\ref{MpLN}). We discuss the results for the
linear and the nonlinear $\sigma$ model, with and without explicitly broken
chiral symmetry. Furthermore, we investigate the counter-term renormalisation
(CTR) method discussed in Appendix \ref{renorm} and the so-called trivial
regularization (TR) where the vacuum contribution of the tadpole integral is
set to zero. This is strictly speaking not an entirely consistent procedure
because these ``vacuum'' contributions actually have an implicit temperature
dependence: they depend on the self-consistently computed particle masses
which are functions of temperature. On the other hand, the CTR method does not
have this shortcoming because the counter terms used to eliminate the
divergences are (infinite) constants independent of temperature.

In the TR method the parameters are determined by solving Eqs.\ (\ref{con}),
(\ref{Msigma}), and (\ref{Mpi}) in the vacuum,%
\begin{equation}
h=m_{\pi}^{2}f_{\pi}\text{ },\text{ \ \ }\dfrac{1}{\varepsilon}=\dfrac
{m_{\sigma}^{2}-m_{\pi}^{2}}{f_{\pi}^{2}}\ ,\text{ \ \ }\upsilon_{0}%
^{2}=f_{\pi}^{2}-2\varepsilon m_{\pi}^{2}\;.\label{param-tr}%
\end{equation}
Similarly, in the CTR method the parameters are obtained from the
solutions of the renormalized equations (\ref{phiren}), (\ref{mpiren}), and
(\ref{msiren}) at $T=0,$%
\begin{align}
h  &  =f_{\pi}\left[  m_{\pi}^{2}+\frac{1}{16\pi^{2}\varepsilon}\,\left(
m_{\sigma}^{2}\ln\frac{m_{\sigma}^{2}}{\mu^{2}}-m_{\sigma}^{2}+\mu^{2}\right)
\right]  \;,\text{ \ \ }\dfrac{1}{\varepsilon}=\dfrac{m_{\sigma}^{2}-m_{\pi
}^{2}}{f_{\pi}^{2}}\ ,\nonumber\\
\upsilon_{0}^{2}  &  =f_{\pi}^{2}-2\varepsilon m_{\pi}^{2}+\frac{1}{16\pi^{2}%
}\,\left[  m_{\sigma}^{2}\,\ln\frac{m_{\sigma}^{2}}{\mu^{2}}-m_{\sigma}%
^{2}+\mu^{2}+3\left(m_{\pi}^{2}\,\ln\frac{m_{\pi}^{2}}{\mu^{2}}-m_{\pi}^{2}%
+\mu^{2}\right) \right] \;. \label{param-ctr}%
\end{align}
Note that, in the chiral limit, $h \rightarrow 0^+$, where $m_\pi
\rightarrow 0$, the first equation requires to choose the
renormalization scale $\mu = m_\sigma$.
\begin{figure}
[ptb]
\begin{center}
\includegraphics[
height=2.38in,
width=5.5097in
]%
{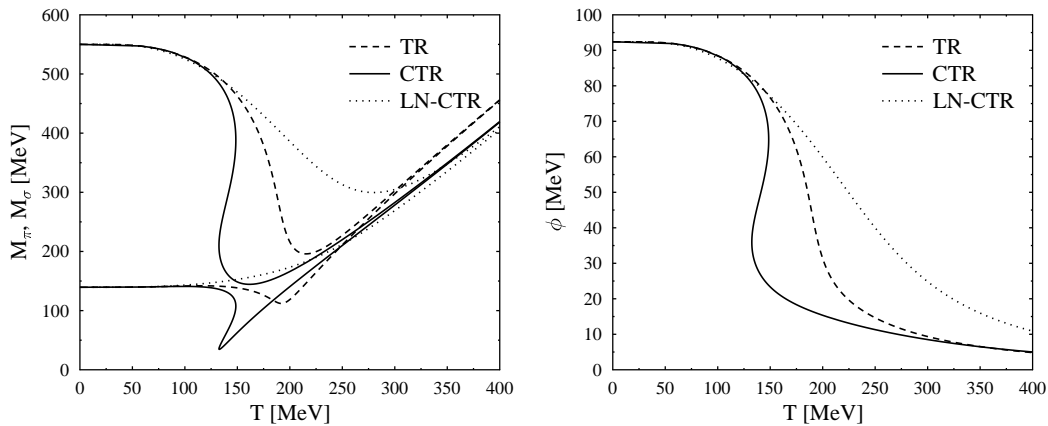}%
\caption{The pion mass, the sigma mass, and the condensate as a function of
$T$ in the $O(4)$ linear model in case of explicitly broken symmetry for
$m_{\sigma}=550$ MeV and different renormalization schemes.}%
\label{lin-550}%
\end{center}
\end{figure}

\subsection{Linear model with explicitly broken symmetry}

In Fig.\ \ref{lin-ct} we show the masses of the pion and the $\sigma$ meson,
as well as the condensate as a function of temperature for different values of
the vacuum $\sigma$ mass $m_{\sigma}$. One observes that the condensate
decreases as a function of temperature, which is a consequence of the
restoration of chiral symmetry. Depending on the value of $m_{\sigma}$, chiral
symmetry restoration may proceed via a phase transition. In the CTR scheme,
the phase transition is of second order for $m_{\sigma}\simeq500$ MeV, and of
first order for larger values of $m_{\sigma}$. For smaller values, however,
the transition is only crossover. In the chirally restored phase, the
condensate is always nonzero because of the small explicit breaking of chiral
symmetry due to non-vanishing quark masses (which gives rise to a nonzero pion
mass $m_{\pi}=139.5$ MeV). Since the results for the TR method are
qualitatively similar, we do not show them explicitly, but we remark that the
second-order transition occurs for larger values of the vacuum $\sigma$ mass,
$m_{\sigma}\simeq700$ MeV. Note that a crossover transition is also found in
lattice QCD calculations \cite{1}. This, however, does not imply that the mass
of the $\sigma$ meson as the chiral partner of the pion must be small. In
fact, the identification of the chiral partner of the pion is a long-debated
issue, see Refs.\ \cite{amslerrev} and refs.\ therein.

Figure \ref{lin-550} shows the effect of different regularization
resp.\ renormalization schemes, as well as different approximation schemes on
the behavior of the masses and the condensate as functions of temperature. We
keep the vacuum mass of the $\sigma$ meson fixed to $m_{\sigma}=550$ MeV. In
the CTR scheme, the system exhibits a first-order phase transition. When using
the TR method, however, one observes a crossover transition. In the large-$N$
limit with CTR, the chiral transition is always crossover, independent of the
mass of the $\sigma$ meson. In Fig.\ \ref{lin-550}, the crossover transition
is observed to be smoother for the large-$N$ approximation with CTR than for
the other cases. In fact, with this renormalization scheme, the smoothness is
proportional to $m_{\sigma}$. We shall see in the next section that the
transition disappears as we approach the nonlinear limit $m_{\sigma
}\rightarrow\infty$. This, however, does not happen for the TR method.

\subsection{Nonlinear model with explicitly broken symmetry}%

\begin{figure}
[ptb]
\begin{center}
\includegraphics[
height=2.3756in,
width=5.4993in
]%
{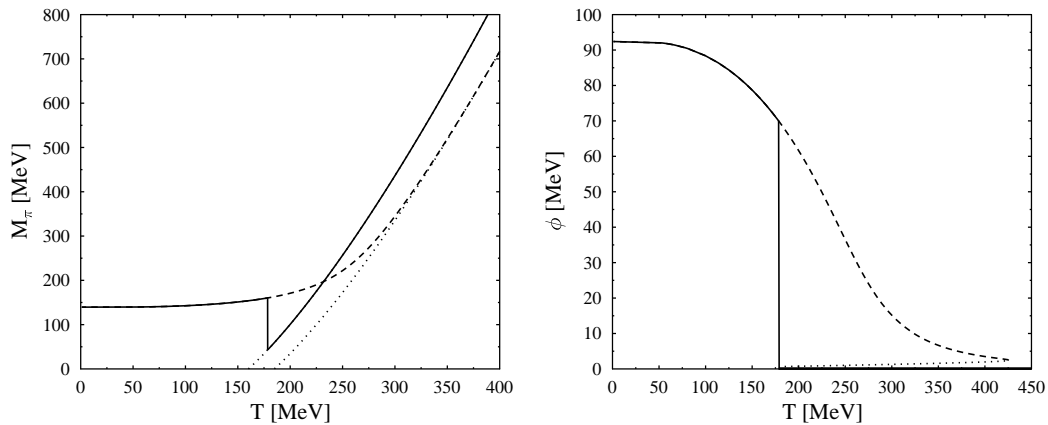}%
\caption{The pion mass and the condensate as a function of $T$ in the $O(4)$
nonlinear model in case of explicitly broken symmetry using the TR-scheme for
$m_{\sigma}\rightarrow\infty$ (in practice $m_{\sigma}=250$ GeV is used).
\ The solid line shows the physical case which corresponds to the
global minimum of the effective potential. The dashed and dotted lines
show the unstable or metastable solution of the gap
equations, which corresponds to the local minimum (dashed) or maximum
(dotted) of the effective potential.}%
\label{h}%
\end{center}
\end{figure}
\begin{figure}
[ptb]
\begin{center}
\includegraphics[
height=2.4025in,
width=5.5616in
]%
{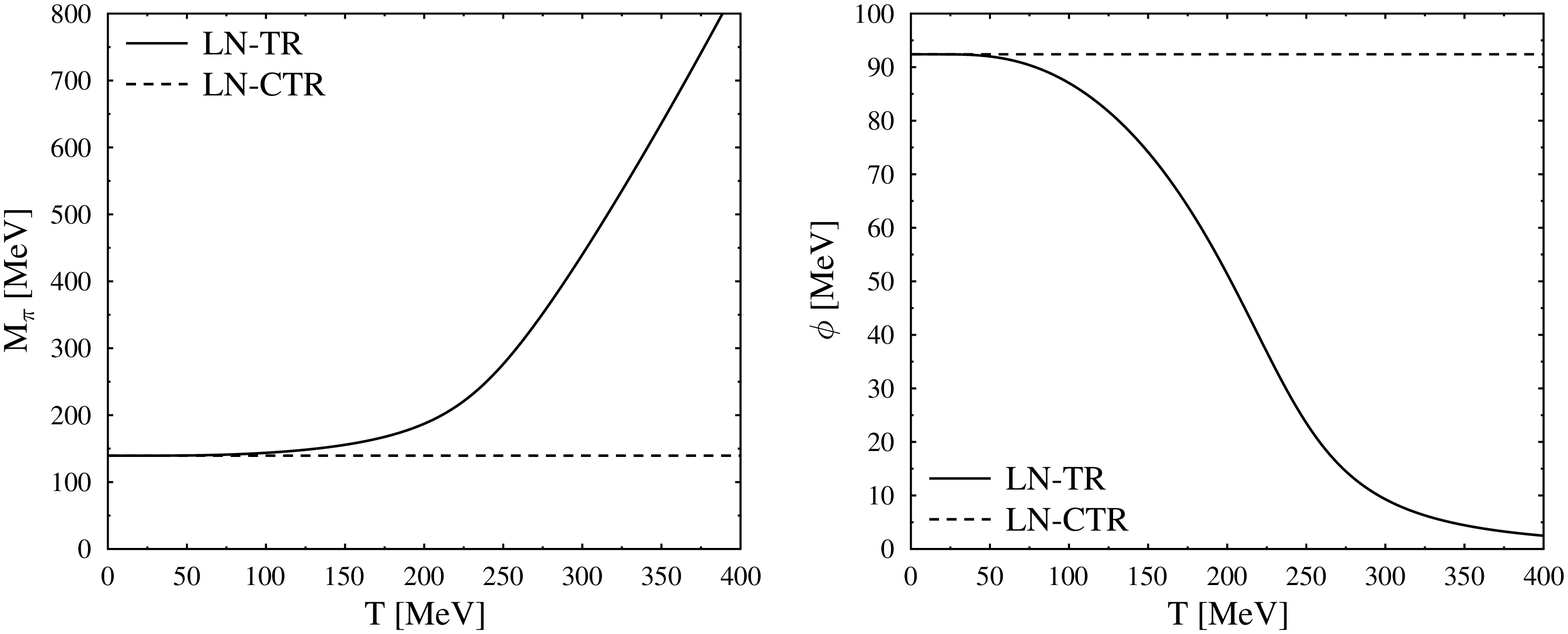}%
\caption{The pion mass and the condensate as a function of $T$ in the $O(4)$
nonlinear model in case of explicitly broken symmetry using the
large-$N$ approximation in the TR-scheme (full) and CTR-scheme (dashed) for
$m_{\sigma}\rightarrow\infty$.}%
\label{nl-ln-h}%
\end{center}
\end{figure}

In the nonlinear model the results are obtained by solving
(the properly renormalized) Eqs.\ (\ref{con}), (\ref{Msigma}), and
(\ref{Mpi}) in the
limit $\varepsilon\rightarrow$ $0^{+}.$ Because of the relation $1/\varepsilon
=\left(  m_{\sigma}^{2}-m_{\pi}^{2}\right)  /f_{\pi}^{2},$
Eqs.\ (\ref{param-tr}) and (\ref{param-ctr}), the nonlinear limit is
equivalent to sending $m_{\sigma}$ to infinity. In this case, when the TR
method is used, the phase transition is of first order, with a rather large
discontinuity in the condensate at a critical temperature of $T_{c}%
\simeq178.6$ MeV, see Fig.\ \ref{h}. The condensate is very small above
$T_{c}$, but still nonzero, because of explicit symmetry breaking. The
first-order nature of the transition is in line with the expectation from the
linear case, where the transition becomes first order when the $\sigma$ mass
is sufficiently large. Below $T_{c}$ the $\sigma$ mass is infinitely heavy and
there are only pionic excitations. Above $T_{c}$ the masses of $\sigma$ meson
and pion become degenerate.

In the large-$N$ limit of the one-loop approximation and with the TR method,
the phase transition is crossover with $T_{c}\simeq185$ MeV, see
Fig.\ \ref{nl-ln-h}. In this case the $\sigma$ field remains infinitely heavy
also above $T_{c}$. This is the main difference to the previous case, where
the $\sigma$ meson becomes degenerate with the pion above $T_{c}$. It is at
first sight surprising that this small difference can cause such a drastic
change in the order of the phase transition. The explanation lies in a
comparison of the equations in the one-loop approximation (\ref{con}),
(\ref{Msigma}), and (\ref{Mpi}) with those in the large-$N$ limit,
Eqs.\ (\ref{gapLN}) -- (\ref{MpLN}). Since the $\sigma$ meson is infinitely
heavy below $T_{c}$, there is no contribution from this mode to these
equations. However, above $T_{c}$, thermal fluctuations of the $\sigma$ meson
can contribute in the one-loop approximation, while they remain absent in the
large-$N$ limit of the one-loop approximation. This is sufficient to drive the
transition to first order in the one-loop approximation.

In the one-loop approximation and using the CTR scheme, the parameter space of
the model does not give physically meaningful solutions in the nonlinear case
$m_{\sigma}\rightarrow\infty$. In this case, $\phi\rightarrow0$ and
$M_{\sigma},$ $M_{\pi}\rightarrow\infty$ for all values of $T$. On the other
hand, the large-$N$ limit of the one-loop approximation allows for a solution,
however, the transition disappears completely, the condensate and the masses
retain their constant tree-level values for all $T>0$: $\phi=f_{\pi},$
$M_{\sigma}=m_{\sigma},$ $M_{\pi}=m_{\pi}$, see Fig.\ \ref{nl-ln-h}.

\subsection{Linear model in the chiral limit}

%

\begin{figure}
[ptb]
\begin{center}
\includegraphics[
height=2.4241in,
width=5.6135in
]%
{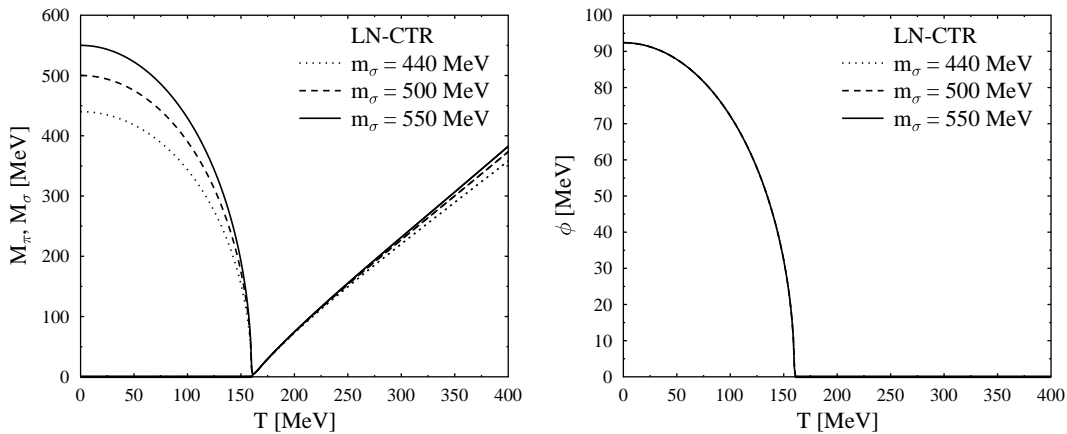}%
\caption{The pion mass, the sigma mass, and the condensate as a
function of $T$ in the large-$N$ limit of the one-loop approximation of
the $O(4)$ linear $\sigma$ model in the chiral limit.}%
\label{ln-cl}%
\end{center}
\end{figure}

The chiral limit is obtained by taking $h\rightarrow0^{+}$. Combining
Eqs.\ (\ref{con}) and (\ref{Mpi}) we see that
\begin{equation}
\phi\left[  M_{\pi}^{2} + \frac{4}{N\varepsilon} \int_{K} G_\sigma(K)\right]  = h
\longrightarrow0^{+}\;,
\end{equation}
which can only be fulfilled if the $\sigma$ tadpole exactly cancels $M_{\pi
}^{2}$. This, however, is only possible, if the pion becomes tachyonic,
$M_{\pi}^{2}<0$, since the thermal as well as the (finite) vacuum contribution
to the tadpole are always positive (semi-)definite. As a consequence, we can
only show results in the large-$N$ limit, since there this problem is absent,
cf.\ Eqs.\ (\ref{gapLN}).

In Fig.\ \ref{ln-cl} we show the behavior of the masses and the condensate as
functions of temperature for various values of the vacuum $\sigma$ mass in the
large-$N$ limit in the CTR scheme (the results for the TR method are
qualitatively similar, therefore we do not show them explicitly). The results
of Fig.\ \ref{ln-cl} are in agreement with universality class arguments which
predict a second-order phase transition. In the phase where chiral symmetry is
spontaneously broken the pions are massless in accordance with Goldstone's
theorem. Above the critical temperature the chiral partners become degenerate
in mass. The condensate as a function of temperature is independent of the
value of $m_{\sigma}$. This can be seen as follows. We subtract
Eq.\ (\ref{MpLN}) at $T=0$ (where $\phi=f_{\pi}$) from the same equation at an
arbitrary temperature $T\leq T_{c}$, where $T_{c}$ is the phase transition
temperature. Since in the phase of broken chiral symmetry we always have
$M_{\pi}\equiv0$, the result is
\begin{equation}
0=\phi^{2}(T)-f_{\pi}^{2}+N\,\frac{T^{2}}{12}\;, \label{phiofT}%
\end{equation}
where the thermal contribution to the tadpole integral could be determined
analytically at all temperatures $T\leq T_{c}$ because $M_{\pi}=0$. The term
$\upsilon_{0}^{2}$, as well as the vacuum contributions to the tadpole
integrals cancel when taking the difference. The critical temperature $T_{c}$
can be easily deduced from Eq.\ (\ref{phiofT}) noting that $\phi(T_{c})=0$.
The result is $T_{c}=\sqrt{12/N}\,f_{\pi}=\sqrt{3}\,f_{\pi}$.

\subsection{Nonlinear model in the chiral limit}

%

\begin{figure}
[ptb]
\begin{center}
\includegraphics[
height=2.412in,
width=5.5824in
]%
{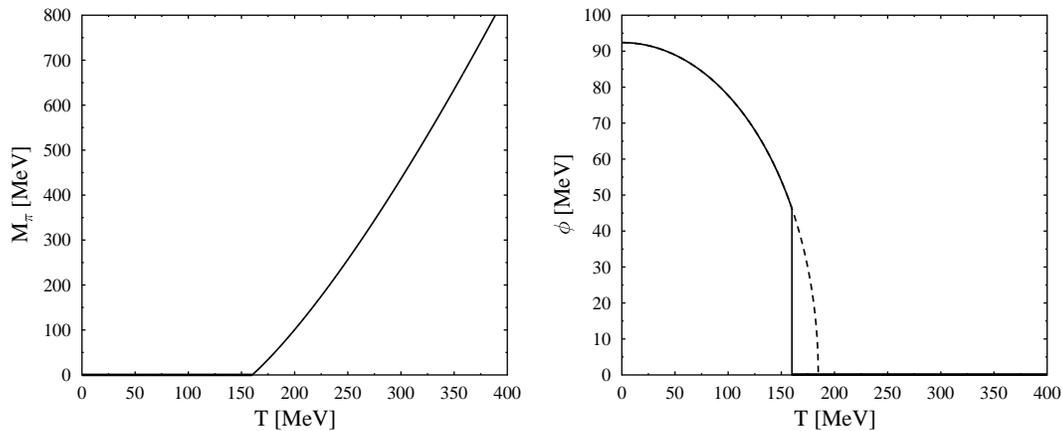}%
\caption{The pion mass and the condensate as a function of $T$ in the $O(4)$
nonlinear model in the chiral limit using the TR-scheme for $m_{\sigma
}\rightarrow\infty$ (in practice $m_{\sigma}=250$ GeV is used). The solid line
corresponds to the physical case. The dashed line shows the metastable
solution of the gap equations which corresponds to the local minimum of the
effective potential.}%
\label{cl}%
\end{center}
\end{figure}
%

\begin{figure}
[ptb]
\begin{center}
\includegraphics[
height=2.4258in,
width=5.6135in
]%
{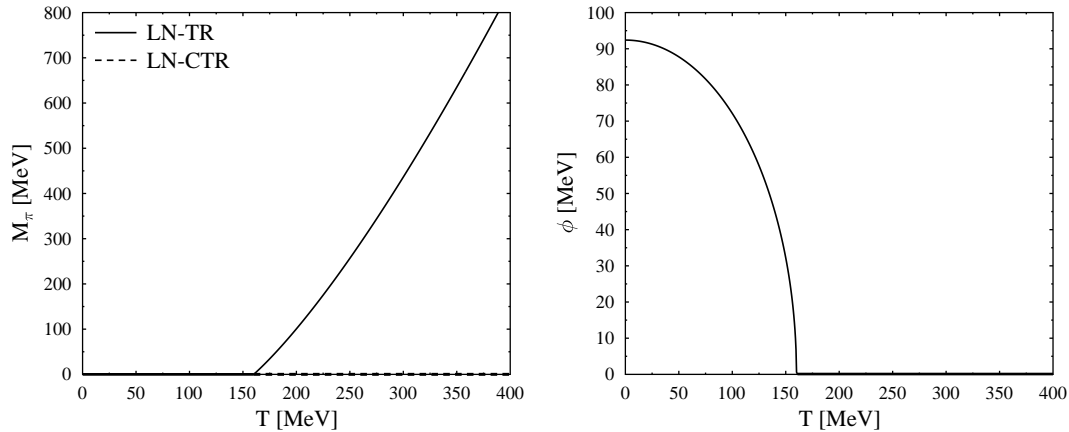}%
\caption{The pion mass and the condensate as a function of $T$ in the $O(4)$
nonlinear model in the chiral limit using the large-$N$ approximation at one-loop
order in the TR-scheme (full) and CTR-scheme (dashed) for $m_{\sigma
}\rightarrow\infty$.}%
\label{nl-ln-cl}%
\end{center}
\end{figure}

In the chiral limit of the nonlinear $O(N)$ model both parameters
$\varepsilon$ and $h$ must be sent to zero. In the one-loop approximation and
in the TR method, pions respect Goldstone's theorem by remaining massless in
the phase of spontaneously broken chiral symmetry, see Fig.\ \ref{cl}. In this
phase, the $\sigma$ field is effectively frozen out due to its infinite mass.
There is a first-order phase transition at a critical temperature $T_{c}%
=\sqrt{3}f_{\pi}$. At this temperature, the condensate drops to zero
discontinuously, while the pion mass starts to increase continuously from zero
above this temperature. In the restored phase, the $\sigma$ meson becomes
degenerate in mass with the pions. This is the reason why $T_{c}$ assumes the
same value as in the large-$N$ limit of the linear model. When inspecting
Eqs.\ (\ref{Msigma}) and (\ref{Mpi}), we observe that they become identical
with Eqs.\ (\ref{MsLN}) and (\ref{MpLN}) for $M_{\sigma}=M_{\pi}=0$ in the
chiral limit and above $T_{c}$ (where $\phi=0$). Therefore, we obtain the same
equation (\ref{phiofT}) that determines the value of $T_{c}$ as in the linear
case in the large-$N$ limit.

However, in the one-loop approximation in the CTR scheme no physical solutions
can be obtained: the condensate goes to zero, $\phi\rightarrow0$, and the
masses of $\sigma$ meson and pion go to infinity, $M_{\sigma}$, $M_{\pi
}\rightarrow\infty$. This situation is similar to the nonlinear case with
explicit symmetry breaking.

In the large-$N$ limit, the phase transition is of second order with a
critical temperature $T_{c}=\sqrt{3}f_{\pi}$, both in the TR method and in the
CTR scheme, see Fig.\ \ref{nl-ln-cl}. Below $T_{c}$ the $\sigma$ mass is
infinite, while the pions are massless, respecting Goldstone's theorem. Above
the critical temperature the masses of the chiral partners become degenerate,
$M_{\sigma}=M_{\pi}>0$ in the TR scheme, and $M_{\sigma}=M_{\pi}=0$ in the CTR
scheme. At first sight, it is surprising that the $\sigma$ field becomes
massless above $T_{c}$. This behavior can be traced to our choice of the
renormalization scale $\mu=m_{\sigma}\rightarrow\infty$. In fact,
this is similar to what was observed in Ref.\ \cite{dirk} (cf.\ Fig.\ 3 of
that work), when increasing the renormalization scale in the large-$N$ limit
in the CTR scheme.

\section{Conclusions}%

\begin{table}[tbp] \centering
\hspace*{-1cm}
\begin{tabular}
[c]{c|c|c||c|c||c|c||c|c|}\cline{2-9}%
$%
\begin{array}
[c]{c}%
\\
\\
\end{array}
$ & $\text{CTR}$ & $\text{CTR}$ & $\text{TR}$ & $\text{TR}$ & $\text{ LN-CTR}$
& $\text{LN-CTR}$ & $\text{LN-TR}$ & $\text{LN-TR}$\\\cline{2-9}%
$%
\begin{array}
[c]{c}%
\\
\\
\end{array}
$ & $m_{\pi}=m_{\pi}^{phys}$ & $m_{\pi}\rightarrow0^{+}$ & $m_{\pi}=m_{\pi
}^{phys}$ & $m_{\pi}\rightarrow0^{+}$ & $m_{\pi}=m_{\pi}^{phys}$ & $m_{\pi
}\rightarrow0^{+}$ & $m_{\pi}=m_{\pi}^{phys}$ & $m_{\pi}\rightarrow0^{+}%
$\\\hline
\multicolumn{1}{|c|}{$\text{lin}$} & $%
\begin{array}
[c]{l}%
\text{second order at}\\
\multicolumn{1}{c}{m_{\sigma}\simeq500\text{ MeV}}%
\end{array}
$ & ${\otimes}$ & $%
\begin{array}
[c]{l}%
\text{second order at}\\
\multicolumn{1}{c}{m_{\sigma}\simeq750\text{ MeV}}%
\end{array}
$ & ${\otimes}$ & crossover & $%
\begin{array}
[c]{l}%
\text{second order}\\
\multicolumn{1}{c}{\text{ }T_{c}=\sqrt{\dfrac{12}{N}}f_{\pi}}%
\end{array}
$ & crossover & $%
\begin{array}
[c]{l}%
\text{second order}\\
\multicolumn{1}{c}{\text{ }T_{c}=\sqrt{\dfrac{12}{N}}f_{\pi}}%
\end{array}
$\\\hline
\multicolumn{1}{|c|}{$\text{nonlinear}$} & ${\otimes}$ & ${\otimes}$ & first
order & $%
\begin{array}
[c]{l}%
\text{first order}\\
\multicolumn{1}{c}{T_{c}=\sqrt{\dfrac{12}{N}}f_{\pi}}%
\end{array}
$ & no transition & \multicolumn{1}{|l||}{$%
\begin{array}
[c]{l}%
\text{second order}\\
\multicolumn{1}{c}{\text{ }T_{c}=\sqrt{\dfrac{12}{N}}f_{\pi}}%
\end{array}
$} & crossover & \multicolumn{1}{|l|}{$%
\begin{array}
[c]{l}%
\text{second order}\\
\multicolumn{1}{c}{\text{ }T_{c}=\sqrt{\dfrac{12}{N}}f_{\pi}}%
\end{array}
$}\\\hline
\end{tabular}%
\caption{Summary of cases studied in this paper. The symbol $\otimes
$ indicates that no reasonable result can be obtained due to tachyonic pion propagation. In those cases the phase transition becomes cross-over for smaller sigma masses and of first order for sigma masses higher than the shown values. $m_{\pi
}=m_{\pi}^{phys}%
$ corresponds to the physical case of nonzero quark masses, $m_{\pi}%
^{phys}=139.5$ MeV. }%
\label{Tab1}%
\end{table}%

In this work we have investigated the linear and the nonlinear $O(N)$ model at
nonzero temperature. An auxiliary field has been used to derive the effective
potential. This method allowed us to establish a simple and mathematically
rigorous relation between the linear and nonlinear versions of the model. This
also leads to differences when comparing our results with previous treatments
of the $O(N)$ model, see below. The equations for the temperature-dependent
masses and the condensate were derived using the CJT formalism. We explicitly
showed that, up to two-loop order, the auxiliary-field method is equivalent to
the standard $O(N)$ linear $\sigma$ model, once the one- and two-point
functions involving the auxiliary field are replaced by their stationary
values. In order to regularize the divergent vacuum terms we applied the
counter-term (CTR) scheme as well as the so-called trivial regularization (TR)
method where divergent terms are simply ignored.

Table \ref{Tab1} shows a compilation of the results for the various scenarios
studied in this paper. The first row summarizes the results for the linear
case, while the second those for the nonlinear case. In the first four columns
we show the results for the one-loop approximation, the first two for the CTR
scheme and the next two for the TR method, for the case of explicit chiral
symmetry breaking and in the chiral limit. The last four columns show the
corresponding results for the large-$N$ limit of the one-loop approximation.
In the cases indicated with a $\otimes$, we were not able to find physically
acceptable solutions due to tachyonic pion propagation. In all other cases, we
indicated the nature of the phase transition and, if independent of the
$\sigma$ mass, the critical temperature. As one observes, $T_{c}=\sqrt
{12/N}\,f_{\pi}\equiv\sqrt{3}\,f_{\pi}$ in the chiral limit for all scenarios,
independent of the details (linear vs.\ nonlinear, or CTR vs.\ TR, or one-loop
approximation vs.\ large-$N$ limit). In the cases where the order of the
transition depends on the $\sigma$ mass, we indicated the value of $m_{\sigma
}$ where the transition is of second order; it is crossover for smaller and of
first order for larger values of $m_{\sigma}$.

We now compare our results to previous work. In Ref.\ \cite{dirk}, the $O(N)$
model for $N=4$ was studied in the CJT formalism without using the
auxiliary-field method. Although not studied in that work, we repeated the
respective calculations varying the $\sigma$ mass. We find that, in the
Hartree-Fock approximation (erroneously named ``Hartree approximation'' in
that paper) and in the case of explicitly broken chiral symmetry, the phase
transition changes from crossover to first order for $m_{\sigma}\simeq940$ MeV
in the TR method and for $m_{\sigma}\simeq680$ MeV in the CTR scheme. This is
consistent with our results obtained with the auxiliary-field method, although
the critical values for $m_{\sigma}$ are somewhat larger for the method of
Ref.\ \cite{dirk}. In the chiral limit, the method of Ref.\ \cite{dirk} yields
a first-order phase transition for all $m_{\sigma}$ values. Furthermore,
Goldstone's theorem is not fulfilled due to a non-vanishing pion mass in the
phase of broken chiral symmetry. In the large-$N$ limit, the results of
Ref.\ \cite{dirk} coincide with ours, since the effective potentials are identical.

The auxiliary-field method has been applied previously to examine properties
of the $O(N)$ model to leading \cite{bochkarev,meyer} and next-to-leading
order in the $1/N$ expansion \cite{warringa,brauner}. To leading order in
$1/N$ the $\sigma$ and $\pi$ fields have the same mass irrespective of whether
chiral symmetry is explicitly or only spontaneously broken. Thus, in the
chiral limit there are four instead of three massless bosons. The phase
transition is of second order with a critical temperature of $T_{c}%
=\sqrt{12/N}\,f_{\pi}.$ In the case of explicitly broken symmetry there is a
crossover phase transition and four massive particles. The key difference in
our study to the afore mentioned Refs.\ \cite{bochkarev,meyer} is the correct
treatment of the limiting process regarding the constraint imposed by the
nonlinearity: the $\sigma$ mass is therefore infinite in the phase of broken
symmetry. To next-to-leading order including renormalization \cite{warringa}
the results change as follows: in the chiral limit there are three Goldstone
bosons since the $\sigma$ field becomes massive. The phase transition is of
second (or higher) order. In the weak-coupling limit the critical temperature
is $T_{c}=\sqrt{12/(N+2)}f_{\pi}$ and above the critical temperature the
masses of the chiral partners become degenerate.

A natural next step is the extension to nonzero chemical potentials
\cite{brauner}. A further interesting study would be the inclusion of
additional scalar singlet states \cite{achim}. Finally, the application of the
auxiliary-field method should also be instructive for more complicated systems
incorporating additional vector and axial vector mesonic degrees of freedom
\cite{stefan}.\newline

\textbf{Acknowledgement}

The authors thank T.\ Brauner, M.\ Grahl, A.\ Heinz, S.\ Leupold, and
H.\ Warringa for interesting discussions. The work of E.\ Seel was supported
by the Helmholtz Research School \textquotedblleft
H-QM\textquotedblright.
We thank the referee for valuable comments which lead to a substantial
modification of an earlier version of this manuscript, and in
particular to the calculations presented in Sec.\ \ref{recover}
and Appendix \ref{casei}.

\appendix

\section{CJT effective potential with non-diagonal propagator}

\label{casei}

In this appendix, we discuss the CJT effective potential for the case where we
do not perform a shift of the $\alpha$ field (denoted as case (i) in
Sec.\ \ref{CJT}). Due to the appearance of non-diagonal progators which mix
the $\alpha$ and $\sigma$ fields, this is more complicated than the case
discussed in the main part of the paper.

\subsection{Tree-level propagators, and vertices}

The starting point is the Lagrangian (\ref{lag2a}), with the tree-level
potential (\ref{U}). From this, we immediately deduce the tree-level
propagator matrix as
\begin{equation}
D^{-1}(K;\phi,\alpha_{0})=\left(
\begin{array}
[c]{cccc}%
D_{\alpha\alpha}^{-1} & D_{\alpha\sigma}^{-1}(\phi) & 0 & \cdots\\[0.1cm]%
D_{\sigma\alpha}^{-1}(\phi) & D_{\sigma\sigma}^{-1}(K;\alpha_{0}) & 0 &
\cdots\\[0.1cm]%
0 & 0 & D_{\pi\pi}^{-1}(K;\alpha_{0}) & \\
\vdots & \vdots &  & \ddots
\end{array}
\right)  =\left(
\begin{array}
[c]{cccc}%
\displaystyle\frac{N\varepsilon}{4} & i\phi & 0 & \cdots\\
i\phi & -K^{2}+i\alpha_{0} & 0 & \cdots\\
0 & 0 & -K^{2}+i\alpha_{0} & \\
\vdots & \vdots &  & \ddots
\end{array}
\right)  \;. \label{Dinvmatrix}%
\end{equation}
Note the following relations between the inverse tree-level propagators in the
shifted, Eq.\ (\ref{pro-alpha}), and unshifted, Eq.\ (\ref{Dinvmatrix}),
cases: $\bar{D}_{\alpha}^{-1}\equiv D_{\alpha\alpha}^{-1}$ and $\bar{D}_{\pi
}^{-1}(K;\alpha_{0})\equiv D_{\pi\pi}^{-1}(K;\alpha_{0})$, while $\bar
{D}_{\sigma}^{-1}(K;\phi,\alpha_{0})=D_{\sigma\sigma}^{-1}(K;\alpha_{0}%
)+4\phi^{2}/(N\varepsilon)$.

The Lagrangian (\ref{lag2a}) contains only two three-point tree-level
vertices, where one $\alpha$ field interacts with either two $\sigma$ or two
{\boldmath$\mathbf{\pi}$} fields, respectively. These are the same vertices
that also appear in case (ii), see Sec.\ \ref{IIIA}.

\subsection{CJT effective potential}

The CJT effective potential assumes the form
\begin{equation}
V_{\mathrm{eff}}(\phi,\alpha_{0},G)=U(\phi,\alpha_{0})+\dfrac{1}{2}\int
_{K}\mathrm{Tr}\,\left[  \ln G^{-1}(K)+D^{-1}(K;\phi,\alpha_{0}%
)\,G(K)-1\right]  +V_{2}(G)\;, \label{veff}%
\end{equation}
where the two-point function $G(K)$ is an $(N+1)\times(N+1)$--matrix, just
like the inverse tree-level propagator $D^{-1}(K;\phi,\alpha_{0})$ in
Eq.\ (\ref{Dinvmatrix}). The term $V_{2}(G)$ represents the sum of all
two-particle irreducible diagrams constructed from $G(K)$ and the two
different three-point vertices\ in Eq.\ (\ref{lag2a}) (which do not depend on
the one-point functions $\phi$ and $\alpha_{0}$).

The stationary conditions for the effective potential are given by
\begin{equation}
\dfrac{\delta V_{\mathrm{eff}}}{\delta\phi}=0\text{ },\ \ \dfrac{\delta
V_{\mathrm{eff}}}{\delta\alpha_{0}}=0\text{ },\text{ }\dfrac{\delta
V_{\mathrm{eff}}}{\delta G_{ij}(K)}=0\text{ },\;\;\;\;i,j=\alpha
\,,\;\sigma\,,\;\pi_{1}\,,\;\ldots\,,\;\pi_{N-1}\;. \label{nlsm-st-cond-a}%
\end{equation}
This leads to the following equations for the two condensates $\phi$ and
$\alpha_{0}$
\begin{align}
h  &  =i\alpha_{0}\phi+\frac{i}{2}\int_{K}\left[  G_{\sigma\alpha
}(K)+G_{\alpha\sigma}(K)\right]  \text{ },\label{con1a}\\
i\alpha_{0}  &  =\dfrac{2}{N\varepsilon}\left[  \phi^{2}-\upsilon_{0}^{2}%
+\int_{K}G_{\sigma\sigma}(K)+(N-1)\int_{K}G_{\pi\pi}(K)\right]  \text{ . }
\label{a1a}%
\end{align}
The equation for $\phi$ is now different from case (ii), see Eq.\ (\ref{con1}%
), but the equation for $\alpha_{0}$ remains the same, cf.\ Eq.\ (\ref{a1}).
The two-point function has the matrix elements
\begin{equation}
G_{ji}^{-1}(K)=D_{ji}^{-1}(K;\phi,\alpha_{0})+\Pi_{ji}(K)\;, \label{m2a}%
\end{equation}
where the one-particle irreducible (1PI) self-energy is
\begin{equation}
\Pi_{ji}(K)=2\,\frac{\delta V_{2}(G)}{\delta G_{ij}(K)}\;,\;\;\;\;i,j=\alpha
\,,\;\sigma\,,\;\pi_{1}\,,\;\ldots\,,\;\pi_{N-1}\;. \label{1PI0}%
\end{equation}

It is instructive to formally invert the full inverse two-point function
$G^{-1}$ in order to obtain the full two-point function $G$. From the Dyson
equation (\ref{m2a}) we observe that $G^{-1}$ has a similar matrix structure
as the inverse tree-level propagator (\ref{Dinvmatrix}). We assume that
inverting $G^{-1}$ preserves this structure, i.e.,
\begin{equation}
G=\left(
\begin{array}
[c]{cccc}%
G_{\alpha\alpha} & G_{\alpha\sigma} & 0 & \cdots\\[0.1cm]%
G_{\sigma\alpha} & G_{\sigma\sigma} & 0 & \cdots\\[0.1cm]%
0 & 0 & G_{\pi\pi} & \\
\vdots & \vdots &  & \ddots
\end{array}
\right)  \;. \label{Gmatrix}%
\end{equation}
Obviously, $G_{\pi\pi}=(G_{\pi\pi}^{-1})^{-1}$. However, inverting the
$2\times2$ matrix corresponding to the $\alpha-\sigma$ sector is more
complicated. From the condition
\begin{equation}
\left(
\begin{array}
[c]{cc}%
1 & 0\\
0 & 1
\end{array}
\right)  =\left(
\begin{array}
[c]{cc}%
G_{\alpha\alpha}^{-1} & G_{\alpha\sigma}^{-1}\\
G_{\sigma\alpha}^{-1} & G_{\sigma\sigma}^{-1}%
\end{array}
\right)  \left(
\begin{array}
[c]{cc}%
G_{\alpha\alpha} & G_{\alpha\sigma}\\
G_{\sigma\alpha} & G_{\sigma\sigma}%
\end{array}
\right)
\end{equation}
we obtain
\begin{align}
G_{\alpha\alpha}  &  =\left[  G_{\alpha\alpha}^{-1}-G_{\alpha\sigma}%
^{-1}\,\frac{1}{G_{\sigma\sigma}^{-1}}\,G_{\sigma\alpha}^{-1}\right]
^{-1}\;,\nonumber\\
G_{\sigma\sigma}  &  =\left[  G_{\sigma\sigma}^{-1}-G_{\sigma\alpha}%
^{-1}\,\frac{1}{G_{\alpha\alpha}^{-1}}\,G_{\alpha\sigma}^{-1}\right]
^{-1}\;,\nonumber\\
G_{\alpha\sigma}  &  =-\frac{1}{G_{\alpha\alpha}^{-1}}\,G_{\alpha\sigma}%
^{-1}G_{\sigma\sigma}=\left[  G_{\sigma\alpha}^{-1}-G_{\sigma\sigma}^{-1}%
\frac{1}{G_{\alpha\sigma}^{-1}}G_{\alpha\alpha}^{-1}\right]  ^{-1}%
\;,\nonumber\\
G_{\sigma\alpha}  &  =-\frac{1}{G_{\sigma\sigma}^{-1}}\,G_{\sigma\alpha}%
^{-1}G_{\alpha\alpha}=\left[  G_{\alpha\sigma}^{-1}-G_{\alpha\alpha}^{-1}%
\frac{1}{G_{\sigma\alpha}^{-1}}G_{\sigma\sigma}^{-1}\right]  ^{-1}\;.
\label{Gmatrix2}%
\end{align}
The second equalities in the last two equations follow by inserting the
explicit expressions for $G_{\sigma\sigma}$ and $G_{\alpha\alpha}$ from the
first two equations. If we assume that $\Pi_{\sigma\alpha}=\Pi_{\alpha\sigma}%
$, then Eq.\ (\ref{m2a}) implies that $G_{\sigma\alpha}^{-1}=G_{\alpha\sigma
}^{-1}$ at the stationary point of $V_{\mathrm{eff}}$. Since $G_{\sigma\sigma
}^{-1}$ and $G_{\alpha\alpha}^{-1}$ are purely numbers, from the last two
equations (\ref{Gmatrix2}) we then obtain $G_{\sigma\alpha}=G_{\alpha\sigma}$.
On the other hand, if we assume the latter, then, from Eqs.\ (\ref{Gmatrix2}),
we conclude that $G_{\sigma\alpha}^{-1}=G_{\alpha\sigma}^{-1}$, from which we
immediately conclude via Eq.\ (\ref{m2a}) that $\Pi_{\sigma\alpha}=\Pi
_{\alpha\sigma}$. In the following, we will therefore make frequent use of the
symmetry property $G_{\sigma\alpha}=G_{\alpha\sigma}$.

With the explicit form (\ref{Gmatrix2}), we can rewrite the one-loop terms in
the effective potential (\ref{veff}). For the first term we obtain
\begin{align}
\mathrm{Tr}\,\ln\,G^{-1}  &  \equiv\ln\,\mathrm{det}\,G^{-1}=\ln
\,\mathrm{det}\left(
\begin{array}
[c]{cc}%
G_{\alpha\alpha}^{-1} & G_{\alpha\sigma}^{-1}\\
G_{\sigma\alpha}^{-1} & G_{\sigma\sigma}^{-1}%
\end{array}
\right)  +(N-1)\,\ln\,G_{\pi\pi}^{-1}\nonumber\\
&  =\ln\left[  G_{\alpha\alpha}^{-1}G_{\sigma\sigma}^{-1}-G_{\alpha\sigma
}^{-1}G_{\sigma\alpha}^{-1}\right]  +(N-1)\,\ln\,G_{\pi\pi}^{-1}\nonumber\\
&  =\ln G_{\alpha\alpha}^{-1}+\ln\left[  G_{\sigma\sigma}^{-1}-\frac
{1}{G_{\alpha\alpha}^{-1}}G_{\alpha\sigma}^{-1}G_{\sigma\alpha}^{-1}\right]
+(N-1)\,\ln\,G_{\pi\pi}^{-1}\nonumber\\
&  =\ln G_{\alpha\alpha}^{-1}+\ln[G_{\sigma\sigma}]^{-1}+(N-1)\,\ln\,G_{\pi
\pi}^{-1}\;, \label{1loop1}%
\end{align}
where the last equality follows from comparison with the second equation
(\ref{Gmatrix2}). Note that $[G_{\sigma\sigma}]^{-1}=G_{\sigma\sigma}%
^{-1}-G_{\alpha\sigma}^{-1}G_{\sigma\alpha}^{-1}/G_{\alpha\alpha}^{-1}\neq
G_{\sigma\sigma}^{-1}$. To make the notation unambiguous, we put brackets
around $G_{\sigma\sigma}$ before inversion. For the second one-loop term we
compute with the help of Eqs.\ (\ref{Dinvmatrix}) and (\ref{Gmatrix})
\begin{equation}
\mathrm{Tr}\left[  D^{-1}G\right]  =D_{\alpha\alpha}^{-1}G_{\alpha\alpha
}+D_{\alpha\sigma}^{-1}G_{\sigma\alpha}+D_{\sigma\alpha}^{-1}G_{\alpha\sigma
}+D_{\sigma\sigma}^{-1}G_{\sigma\sigma}+(N-1)D_{\pi}^{-1}G_{\pi\pi}\;.
\label{1loop2}%
\end{equation}
Inserting Eqs.\ (\ref{1loop1}) and (\ref{1loop2}) into Eq.\ (\ref{veff}), we
obtain
\begin{align}
V_{\mathrm{eff}}(\phi,\alpha_{0},G)  &  =U(\phi,\alpha_{0})+\frac{1}{2}%
\int_{K}\left[  \ln G_{\alpha\alpha}^{-1}(K)+\ln[G_{\sigma\sigma}%
(K)]^{-1}+(N-1)\ln G_{\pi\pi}^{-1}(K)\right] \nonumber\\
&  +\frac{1}{2}\int_{K}\left[  D_{\alpha\alpha}^{-1}G_{\alpha\alpha
}(K)+D_{\alpha\sigma}^{-1}(\phi)G_{\sigma\alpha}(K)+D_{\sigma\alpha}^{-1}%
(\phi)G_{\alpha\sigma}(K)+D_{\sigma\sigma}^{-1}(K;\alpha_{0})G_{\sigma\sigma
}(K)\right. \nonumber\\
&  \hspace*{0.7cm}+\left.  (N-1)D_{\pi\pi}^{-1}(K;\alpha_{0})G_{\pi\pi
}(K)-(N+1)\right]  +V_{2}(G)\;. \label{veff2}%
\end{align}

\subsection{One-loop approximation}

In one-loop approximation, $V_{2}(G)\equiv0,$ Eqs.\ (\ref{con1a}) and
(\ref{a1a}) for the condensates $\phi$ and $\alpha_{0}$ remain unchanged. For
vanishing $V_{2}(G)$ the 1PI self-energy is equal to zero, $\Pi_{ji}(K)=0,$ and
\begin{equation}
G_{ji}^{-1}(K)=D_{ji}^{-1}(K;\phi,\alpha_{0})\;,\;\;\;\;i,j=\alpha
\,,\;\sigma\,,\;\pi_{1}\,,\;\ldots\,,\;\pi_{N-1}\;.
\end{equation}
The full two-point functions (\ref{Gmatrix2}) then become
\begin{align}
G_{\sigma\sigma}(K)  &  =\left[  D_{\sigma\sigma}^{-1}(K;\alpha_{0}%
)-\frac{D_{\sigma\alpha}^{-1}(\phi)D_{\alpha\sigma}^{-1}(\phi)}{D_{\alpha
\alpha}^{-1}}\right]  ^{-1}=\left(  -K^{2}+i\alpha_{0}+\frac{4\phi^{2}%
}{N\varepsilon}\right)  ^{-1}\;,\nonumber\\
G_{\alpha\alpha}(K)  &  =\left[  D_{\alpha\alpha}^{-1}-\frac{D_{\alpha\sigma
}^{-1}(\phi)D_{\sigma\alpha}^{-1}(\phi)}{D_{\sigma\sigma}^{-1}(K;\alpha_{0}%
)}\right]  ^{-1}=\left(  \frac{N\varepsilon}{4}+\frac{\phi^{2}}{-K^{2}%
+i\alpha_{0}}\right)  ^{-1}=\frac{4}{N\varepsilon}\left[  1-\frac{4\phi^{2}%
}{N\varepsilon}\,G_{\sigma\sigma}(K)\right]  \;,\nonumber\\
G_{\alpha\sigma}(K)  &  =-\frac{4i\phi}{N\varepsilon}\,G_{\sigma\sigma
}(K)\equiv G_{\sigma\alpha}(K)\;, \label{Gmatrix1loopcasei}%
\end{align}
where the symmetry of the mixed two-point function, $G_{\sigma\alpha
}=G_{\alpha\sigma}$ is automatic. The two-point function for the pion simply
reads
\begin{equation}
G_{\pi\pi}(K)=\left(  -K^{2}+i\alpha_{0}\right)  ^{-1}\;.
\end{equation}
The $\sigma$ and pion two-point functions can be written in the form
\begin{equation}
G_{\sigma\sigma}(K)=(-K^{2}+M_{\sigma}^{2})^{-1}\;,\;\;\;\;G_{\pi\pi
}(K)=(-K^{2}+M_{\pi}^{2})^{-1}\;, \label{Gnondiag}%
\end{equation}
with the same mass parameters as in Eqs.\ (\ref{Msigma}) and (\ref{Mpi}),
since the condensate equation (\ref{a1a}) for $\alpha_{0}$ is identical to the
one in the shifted case, Eq.\ (\ref{a1}).

Substituting $i\alpha_{0}$ by Eq.\ (\ref{a1a}), the condensate equation
(\ref{con1a}) becomes
\begin{equation}
h=i\alpha_{0}\phi+\frac{4\phi}{N\varepsilon}\int_{K}G_{\sigma\sigma}%
(K)=\frac{2\phi}{N\varepsilon}\left[  \phi^{2}-\upsilon_{0}^{2}+3\int
_{K}G_{\sigma\sigma}(K)+(N-1)\int_{K}G_{\pi\pi}(K)\right]  \;,
\end{equation}
where we have used Eq.\ (\ref{Gmatrix1loopcasei}) to rewrite $G_{\alpha\sigma
}$ and $G_{\sigma\alpha}$\ in terms of $G_{\sigma\sigma}.$ Since $G_{\sigma}=
G_{\sigma\sigma}$ and $G_{\pi}= G_{\pi\pi}$, this equation is identical with
the condensate equation for $\phi$ in the shifted case, Eq.\ (\ref{con}). We
have therefore proved that the equations for $M_{\sigma}$ and $M_{\pi}$ and
the condensate equation for $\phi$ are identical to the corresponding
equations in the shifted case (ii).

\subsection{Two-loop approximation}%

\begin{figure}
[ptb]
\begin{center}
\includegraphics[
height=1.2263in,
width=4.3915in
]%
{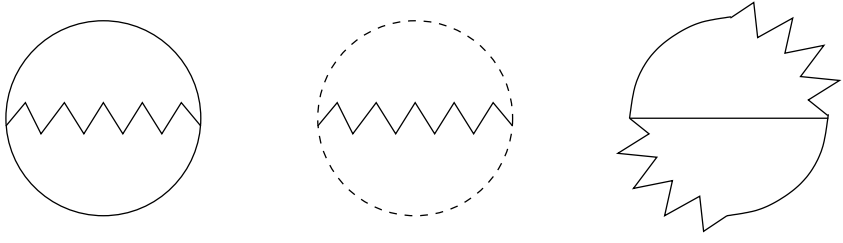}%
\caption{Two-particle irreducible diagrams constructed from the three-point
vertices in Eq.\ (\ref{lag2a}). The full line represents the $\sigma$ field,
the dashed line represents the $\pi$ field and the zigzag line represents the
$\alpha$ field. The non-diagonal propagators $G_{\alpha\sigma}$ and
$G_{\sigma\alpha}$ are denoted by partially full and partially zig-zagged lines.}%
\label{v2(b)}%
\end{center}
\end{figure}

In case (i), to two-loop order there are only the three diagrams of sunset
topology shown in Fig.\ \ref{v2(b)}, resulting in
\begin{equation}
V_{2}(G)=\frac{1}{4}\int_{K}\int_{P}\left\{  G_{\alpha\alpha}(K+P)\left[
G_{\sigma\sigma}(K)G_{\sigma\sigma}(P)+(N-1)G_{\pi\pi}(K)G_{\pi\pi}(P)\right]
+2\,G_{\sigma\sigma}(K+P)G_{\alpha\sigma}(K)G_{\sigma\alpha}(P)\right\}
\text{ .}\label{V2i}%
\end{equation}
Due to the absence of a four-point vertex, there is no two-loop diagram of
double-bubble topology. Comparing Fig.\ \ref{v2(b)} to Fig.\ \ref{V2(a)}, we
notice that there is an additional diagram due to the presence of non-diagonal
propagators, but that the last two diagrams in Fig.\ \ref{V2(a)} are absent,
since there is no vertex proportional to $\phi$.

The equations for the two condensates $\phi$ and $\alpha_{0}$ are again given
by Eqs.\ (\ref{con1a}) and (\ref{a1a}). From Eq.\ (\ref{1PI0}) we immediately
derive the self-energies
\begin{align}
\Pi_{\alpha\alpha}(K)  &  =\frac{1}{2}\int_{P}\left[  G_{\sigma\sigma
}(P)G_{\sigma\sigma}(K-P)+(N-1)G_{\pi\pi}(P)G_{\pi\pi}(K-P)\right]
\;,\nonumber\\
\Pi_{\sigma\alpha}(K)  &  =\int_{P}G_{\sigma\alpha}(P)G_{\sigma\sigma
}(K-P)\;,\nonumber\\
\Pi_{\alpha\sigma}(K)  &  =\int_{P}G_{\alpha\sigma}(P)G_{\sigma\sigma
}(K-P)\;,\nonumber\\
\Pi_{\sigma\sigma}(K)  &  =\int_{P}\left[  G_{\sigma\sigma}(P)G_{\alpha\alpha
}(K-P)+G_{\alpha\sigma}(P)G_{\sigma\alpha}(K-P)\right]  \;,\nonumber\\
\Pi_{\pi\pi}(K)  &  =\int_{P}G_{\pi\pi}(P)G_{\alpha\alpha}(K-P)\;. \label{1PI}%
\end{align}
Since $G_{\alpha\sigma}=G_{\sigma\alpha}$ at the stationary point, we confirm
that $\Pi_{\alpha\sigma}=\Pi_{\sigma\alpha}$.

Replacing $\alpha_{0}$\ by Eq. (\ref{a1a}) the equation for the condensate
$\phi$ reads%
\begin{equation}
\label{gapequ}h=\dfrac{2\phi}{N\varepsilon}\left[  \phi^{2}-\upsilon_{0}%
^{2}+\int_{K}G_{\sigma\sigma}(K)+(N-1)\int_{K}G_{\pi\pi}(K)\right]  +\frac
{i}{2}\int_{K}\left[  G_{\sigma\alpha}(K)+G_{\alpha\sigma}(K)\right]  \text{
}.
\end{equation}
After substituting $\alpha_{0}$\ by Eq. (\ref{a1a}) the Dyson
equations for the full two-point functions are given by%
\begin{align}
G_{\alpha\alpha}^{-1}(K)  &  =D_{\alpha\alpha}^{-1}+\Pi_{\alpha\alpha
}(K)=\frac{N\varepsilon}{4}+\frac{1}{2}\int_{P}\left[  G_{\sigma\sigma
}(P)G_{\sigma\sigma}(K-P)+(N-1)G_{\pi\pi}(P)G_{\pi\pi}(K-P)\right]
\;,\label{Gaa}\\
G_{\sigma\alpha}^{-1}(K)  &  =D_{\sigma\alpha}^{-1}(\phi)+\Pi_{\sigma\alpha
}(K)=i\phi+\int_{P}G_{\sigma\alpha}(P)G_{\sigma\sigma}(K-P)\;,\label{Gsa}\\
G_{\alpha\sigma}^{-1}(K)  &  =D_{\alpha\sigma}^{-1}(\phi)+\Pi_{\alpha\sigma
}(K)=i\phi+\int_{P}G_{\alpha\sigma}(P)G_{\sigma\sigma}(K-P)\;,\label{Gas}\\
G_{\sigma\sigma}^{-1}(K)  &  =D_{\sigma\sigma}^{-1}(K;\alpha_{0})+\Pi
_{\sigma\sigma}(K)=-K^{2}+i\alpha_{0}+\int_{P}\left[  G_{\sigma\sigma
}(P)G_{\alpha\alpha}(K-P)+G_{\alpha\sigma}(P)G_{\sigma\alpha}(K-P)\right]
\nonumber\\
&  =-K^{2}+\dfrac{2}{N\varepsilon}\left[  \phi^{2}-\upsilon_{0}^{2}+\int
_{K}G_{\sigma\sigma}(K)+(N-1)\int_{K}G_{\pi\pi}(K)\right] \nonumber\\
&  +\int_{P}\left[  G_{\sigma\sigma}(P)G_{\alpha\alpha}(K-P)+G_{\alpha\sigma
}(P)G_{\sigma\alpha}(K-P)\right]  \;,\label{Gss}\\
G_{\pi\pi}^{-1}(K)  &  =D_{\pi\pi}^{-1}(K;\alpha_{0})+\Pi_{\pi\pi}%
(K)=-K^{2}+i\alpha_{0}+\int_{P}G_{\pi\pi}(P)G_{\alpha\alpha}(K-P)\;\nonumber\\
&  =-K^{2}+\frac{2}{N\varepsilon}\left[  \phi^{2}-\upsilon_{0}^{2}+\int
_{K}G_{\sigma\sigma}(K)+(N-1)\int_{K}G_{\pi\pi}(K)\right]  +\int_{P}G_{\pi\pi
}(P)G_{\alpha\alpha}(K-P)\;. \label{Gpipi}%
\end{align}

\subsection{Recovering the standard two-loop approximation}

In this subsection, we show that, up to two-loop order, the condensate and
mass equations, the full propagators, as well as the effective potential
become identical with the corresponding quantities for the standard linear
$\sigma$ model, once we eliminate the $\alpha$ field using the condensate
equation (\ref{a1a}), as well as the corresponding propagators at their
stationary values, cf.\ Eqs.\ (\ref{Gaa}) -- (\ref{Gas}).

We first consider Eq.\ (\ref{gapequ}). The $\alpha_{0}$ field has already been
substituted, and we just have to replace $G_{\sigma\alpha}$ and $G_{\alpha
\sigma}$ by their stationary values. To this end, we use $G_{\sigma\alpha} =
G_{\alpha\sigma}$ and the third Eq.\ (\ref{Gmatrix2}), where we substitute
$G_{\alpha\alpha}^{-1}$ and $G_{\alpha\sigma}^{-1}$ from Eqs.\ (\ref{Gaa}) and
(\ref{Gas}). Then, expanding to two-loop order (i.e., retaining only terms of
first order in the self-energies $\Pi_{\alpha\alpha}$ and $\Pi_{\alpha\sigma}%
$),
\begin{align}
\frac{1}{2}\left[  G_{\sigma\alpha}(K) + G_{\alpha\sigma}(K) \right]   &  =
G_{\alpha\sigma}(K) = - \left[  D_{\alpha\alpha}^{-1} + \Pi_{\alpha\alpha
}(K)\right]  ^{-1} \left[  D_{\alpha\sigma}^{-1}(\phi) + \Pi_{\alpha\sigma}(K)
\right]  G_{\sigma\sigma}(K)\nonumber\\
&  \simeq-\, D_{\alpha\alpha} \left[  D_{\alpha\sigma}^{-1} (\phi)
+\Pi_{\alpha\sigma}(K) - D_{\alpha\alpha}\, \Pi_{\alpha\alpha}(K)\,D_{\alpha
\sigma}^{-1} (\phi) \right]  G_{\sigma\sigma}(K)\nonumber\\
&  = - \frac{4i\phi}{N \varepsilon}\, G_{\sigma\sigma}(K) - \frac
{4}{N\varepsilon}\int_{P}G_{\alpha\sigma}(P)G_{\sigma\sigma}(K-P)\nonumber\\
&  + \frac{i \phi}{2}\left(  \frac{4}{N \varepsilon}\right)  ^{2} \int
_{P}\left[  G_{\sigma\sigma}(P)G_{\sigma\sigma}(K-P) +(N-1)G_{\pi\pi}%
(P)G_{\pi\pi}(K-P)\right]  \;.
\end{align}
To two-loop order, we may then replace $G_{\alpha\sigma}(P)$ under the
integral by the first-order contribution $-4i\phi/(N\varepsilon)\,
G_{\sigma\sigma}(P)$. Inserting everything into Eq.\ (\ref{gapequ}), we obtain
Eq.\ (\ref{con12loop}).

Let us now consider the full propagators $G_{\sigma\sigma}$ and $G_{\pi\pi}$.
Since we are working at two-loop order in the effective potential, it is
sufficient to compute these propagators to one-loop order, i.e., by
considering terms up to linear order in the self-energies $\Pi_{ij}(K)$. Thus,
using the stationary values (\ref{Gaa}) -- (\ref{Gss}) we may expand the
(inverse of the) second Eq.\ (\ref{Gmatrix2}) as
\begin{align}
\left[  G_{\sigma\sigma}(K)\right]  ^{-1}  &  = G_{\sigma\sigma}^{-1}(K) -
G_{\sigma\alpha}^{-1}(K)\, \frac{1}{G_{\alpha\alpha}^{-1}(K)}\, G_{\alpha
\sigma}^{-1}(K)\nonumber\\
&  = D_{\sigma\sigma}^{-1}(K;\alpha_{0}) + \Pi_{\sigma\sigma}(K) - \left[
D_{\sigma\alpha}^{-1}(\phi) + \Pi_{\sigma\alpha}(K)\right]  \left[
D_{\alpha\alpha}^{-1} + \Pi_{\alpha\alpha}(K)\right]  ^{-1} \left[
D_{\alpha\sigma}^{-1}(\phi) + \Pi_{\alpha\sigma}(K)\right] \nonumber\\
&  \simeq D_{\sigma\sigma}^{-1}(K;\alpha_{0}) - D_{\alpha\alpha}\,
D_{\sigma\alpha}^{-2}(\phi) +\Pi_{\sigma\sigma}(K) + D_{\alpha\alpha}^{2}
D_{\sigma\alpha}^{-2}(\phi)\, \Pi_{\alpha\alpha}(K) - 2 \, D_{\alpha\alpha}\,
D_{\alpha\sigma}^{-1}(\phi)\, \Pi_{\sigma\alpha}(K)\;, \label{GssZwischen}%
\end{align}
where we have used $D_{\alpha\sigma}^{-1}(\phi) = D_{\sigma\alpha}^{-1}(\phi)$
and $\Pi_{\alpha\sigma}(K) = \Pi_{\sigma\alpha}(K)$. The first two terms are
identical with the tree-level propagator in the shifted case,
\begin{equation}
D_{\sigma\sigma}^{-1}(K;\alpha_{0}) - D_{\alpha\alpha}\, D_{\sigma\alpha}%
^{-2}(\phi) = -K^{2} + i \alpha_{0} + \frac{4 \phi^{2}}{N \varepsilon}
\equiv\bar{D}_{\sigma}^{-1}(K;\phi,\alpha_{0})\;,
\end{equation}
cf.\ Eq.\ (\ref{pro-alpha}). Inserting Eq.\ (\ref{a1a}), this can be written
as
\begin{equation}
\label{Gssa}\bar{D}_{\sigma}^{-1}(K;\phi,\alpha_{0}) = - K^{2} + \dfrac
{2}{N\varepsilon}\left[  3 \phi^{2}-\upsilon_{0}^{2}+\int_{K}G_{\sigma\sigma
}(K)+(N-1)\int_{K}G_{\pi\pi}(K)\right]  \;.
\end{equation}
To one-loop order, i.e., employing the one-loop results
(\ref{Gmatrix1loopcasei}) for the propagators involving the $\alpha$ field,
the remaining terms in Eq.\ (\ref{GssZwischen}) can be written as
\begin{align}
\lefteqn{\Pi_{\sigma\sigma}(K) + D_{\alpha\alpha}^{2} D_{\sigma\alpha}%
^{-2}(\phi)\, \Pi_{\alpha\alpha}(K) - 2 \, D_{\alpha\alpha}\, D_{\alpha\sigma
}^{-1}(\phi)\, \Pi_{\sigma\alpha}(K)}\nonumber\\
&  = \int_{P}\left\{  \frac{}{} G_{\sigma\sigma}(P)G_{\alpha\alpha}(K-P)
+G_{\alpha\sigma}(P)G_{\sigma\alpha}(K-P) \right. \nonumber\\
&  - \left.  \frac{1}{2} \left(  \frac{4 \phi}{N\varepsilon}\right)  ^{2}
\left[  G_{\sigma\sigma}(P)G_{\sigma\sigma}(K-P)+(N-1)G_{\pi\pi}(P)G_{\pi\pi
}(K-P)\right]  - 2 i \frac{4 \phi}{N \varepsilon}G_{\sigma\alpha}%
(P)G_{\sigma\sigma}(K-P)\right\} \nonumber\\
&  \simeq\int_{P}\left\{  \frac{4}{N\varepsilon}\, G_{\sigma\sigma}(P) \left[
1- \frac{4 \phi^{2}}{N \varepsilon}\,G_{\sigma\sigma}(K-P) \right]  - \left(
\frac{4 \phi}{N\varepsilon} \right)  ^{2} G_{\sigma\sigma}(P) G_{\sigma\sigma
}(K-P) \right. \nonumber\\
&  - \left.  \frac{1}{2} \left(  \frac{4 \phi}{N\varepsilon}\right)  ^{2}
\left[  G_{\sigma\sigma}(P)G_{\sigma\sigma}(K-P)+(N-1)G_{\pi\pi}(P)G_{\pi\pi
}(K-P)\right]  - 2 \left(  \frac{4 \phi}{N \varepsilon}\right)  ^{2}%
G_{\sigma\sigma}(P)G_{\sigma\sigma}(K-P)\right\} \nonumber\\
&  = \frac{4}{N\varepsilon} \int_{P} G_{\sigma\sigma}(P) - 2 \left(  \frac{2
\phi}{N\varepsilon} \right)  ^{2} \int_{P} \left[  9\, G_{\sigma\sigma
}(P)G_{\sigma\sigma}(K-P)+(N-1)G_{\pi\pi}(P)G_{\pi\pi}(K-P)\right]  \;.
\label{Gssb}%
\end{align}
Summing Eqs.\ (\ref{Gssa}) and (\ref{Gssb}), we see that $\left[
G_{\sigma\sigma}(K)\right]  ^{-1}$ becomes identical to the full inverse
$\sigma$ propagator in the standard $O(N)$ linear $\sigma$ model,
cf.\ Eq.\ (\ref{Gsigmalsm}). For the inverse pion propagator (\ref{Gpipi}),
we simply have to insert the one-loop result (\ref{Gmatrix1loopcasei}) for
$G_{\alpha\alpha}(K-P)$ in the last term,
\begin{align}
G_{\pi\pi}^{-1}(K)  &  = -K^{2} + \dfrac{2}{N\varepsilon}\left[  \phi
^{2}-\upsilon_{0}^{2}+\int_{K}G_{\sigma\sigma}(K)+(N-1)\int_{K}G_{\pi\pi}(K)
\right]  +\int_{P}G_{\pi\pi}(P)G_{\alpha\alpha}(K-P)\nonumber\\
&  \simeq-K^{2} + \dfrac{2}{N\varepsilon}\left\{  \phi^{2}-\upsilon_{0}%
^{2}+\int_{K}G_{\sigma\sigma}(K)+(N-1)\int_{K}G_{\pi\pi}(K)+ 2 \int_{P}%
G_{\pi\pi}(P) \left[  1 - \frac{4 \phi^{2}}{N\varepsilon}\, G_{\sigma\sigma
}(K-P)\right]  \right\} \nonumber\\
&  = -K^{2} + \dfrac{2}{N\varepsilon}\left\{  \phi^{2}-\upsilon_{0}^{2}%
+\int_{K}G_{\sigma\sigma}(K)+(N+1)\int_{K}G_{\pi\pi}(K) - \frac{8 \phi^{2}%
}{N\varepsilon} \int_{P}G_{\pi\pi}(P) \, G_{\sigma\sigma}(K-P) \right\}  \;.
\end{align}
This is identical with the inverse pion propagator (\ref{Gpilsm}) in the
standard linear $\sigma$ model.

Finally, we show that the two-loop effective potential (\ref{veff2}) becomes
identical with the one for the standard linear $\sigma$ model,
Eq.\ (\ref{v2cjt}), if we replace the expectation value and the full two-point
function for the auxiliary field by their stationary values. We again consider
the tree-level, the one-loop, and the two-loop contributions in
Eq.\ (\ref{Veff}) separately. Since the condensate equation for $\alpha_{0}$
is the same in both cases, cf.\ Eqs.\ (\ref{a1}) and (\ref{a1a}), the
tree-level potential at the stationary value for $\alpha_{0}$ is given by the
same expression as in the shifted case (ii), cf.\ Eq.\ (\ref{treelevel}).

For the one-loop terms, we first prove that, up to two-loop order, the
following identity holds,
\begin{equation}
\label{identity}\ln G_{\alpha\alpha}^{-1} + D_{\alpha\alpha}^{-1}%
G_{\alpha\alpha} + D_{\alpha\sigma}^{-1}G_{\sigma\alpha} + D_{\sigma\alpha
}^{-1}G_{\alpha\sigma} + D_{\sigma\sigma}^{-1}G_{\sigma\sigma} \simeq1 +
\left(  D_{\sigma\sigma}^{-1} - D_{\sigma\alpha}^{-1}\, \frac{1}%
{D_{\alpha\alpha}^{-1}}\, D_{\alpha\sigma}^{-1} \right)  G_{\sigma\sigma} +
const.\;,
\end{equation}
where the last term is a(n irrelevant) constant. Inserting the formal
solutions (\ref{Gmatrix2}) for $G_{\sigma\alpha}$ and $G_{\alpha\sigma}$, the
left-hand side of Eq.\ (\ref{identity}) can be written as
\begin{equation}
\ln\left(  D_{\alpha\alpha}^{-1} +\Pi_{\alpha\alpha}\right)  + \left(
D_{\alpha\alpha}^{-1} - D_{\alpha\sigma}^{-1}\,G_{\sigma\alpha}^{-1}\,
\frac{1}{G_{\sigma\sigma}^{-1}}\right)  G_{\alpha\alpha} + \left(
D_{\sigma\sigma}^{-1} - D_{\sigma\alpha}^{-1}\,G_{\alpha\sigma}^{-1}\,
\frac{1}{G_{\alpha\alpha}^{-1}}\right)  G_{\sigma\sigma}%
\end{equation}
Up to two-loop order, it is sufficient to expand the first term up to first
order in $\Pi_{\alpha\alpha}$,
\begin{equation}
\ln\left(  D_{\alpha\alpha}^{-1} +\Pi_{\alpha\alpha}\right)  \simeq\ln
D_{\alpha\alpha}^{-1} + D_{\alpha\alpha} \,\Pi_{\alpha\alpha}\;.
\end{equation}
Since $\ln D_{\alpha\alpha}^{-1} = \ln N\varepsilon/4$ is an irrelevant
constant, we only need to retain the second term. Using the Dyson equation
(\ref{Gas}) we may then rewrite the left-hand side of Eq.\ (\ref{identity})
as
\begin{align}
\lefteqn{ D_{\alpha\alpha} \,\Pi_{\alpha\alpha} + \left[  D_{\alpha\alpha
}^{-1} - \left(  G_{\alpha\sigma}^{-1} -\Pi_{\alpha\sigma}\right)
\,G_{\sigma\alpha}^{-1}\, \frac{1}{G_{\sigma\sigma}^{-1}}\right]
G_{\alpha\alpha} + \left[  D_{\sigma\sigma}^{-1} - D_{\sigma\alpha}%
^{-1}\,\left(  D_{\alpha\sigma}^{-1} + \Pi_{\alpha\sigma} \right)  \, \frac
{1}{G_{\alpha\alpha}^{-1}}\right]  G_{\sigma\sigma}}\nonumber\\
&  \simeq D_{\alpha\alpha} \,\Pi_{\alpha\alpha} + \left(  G_{\alpha\alpha
}^{-1} - \Pi_{\alpha\alpha} - G_{\alpha\sigma}^{-1}\,G_{\sigma\alpha}^{-1}\,
\frac{1}{G_{\sigma\sigma}^{-1}} + \Pi_{\alpha\sigma}\, G_{\sigma\alpha}^{-1}\,
\frac{1}{G_{\sigma\sigma}^{-1}} \right)  G_{\alpha\alpha}\nonumber\\
&  + \left[  D_{\sigma\sigma}^{-1} - D_{\sigma\alpha}^{-1}\,\left(
D_{\alpha\sigma}^{-1} + \Pi_{\alpha\sigma} \right)  \, \frac{1}{D_{\alpha
\alpha}^{-1}} \left(  1-D_{\alpha\alpha}\, \Pi_{\alpha\alpha} \right)
\right]  G_{\sigma\sigma}\;,
\end{align}
where we have used Eq.\ (\ref{Gaa}) and again expanded up to first order in
$\Pi_{\alpha\alpha}$. The first and the third term in the first parentheses
yield $[G_{\alpha\alpha}]^{-1}$, cf.\ the first Eq.\ (\ref{Gmatrix2}). To
two-loop order, the terms in brackets may be expanded to first order in the
self-energies $\Pi_{ij}$. We then obtain
\begin{equation}
\label{Zwischen3}D_{\alpha\alpha} \,\Pi_{\alpha\alpha} + 1 - \left(
\Pi_{\alpha\alpha} - \Pi_{\alpha\sigma}\, \frac{G_{\sigma\alpha}^{-1}%
}{G_{\sigma\sigma}^{-1}} \right)  G_{\alpha\alpha} + \left(  D_{\sigma\sigma
}^{-1} - \frac{D_{\sigma\alpha}^{-1}\,D_{\alpha\sigma}^{-1}}{D_{\alpha\alpha
}^{-1}} - \frac{D_{\sigma\alpha}^{-1}}{D_{\alpha\alpha}^{-1}}\, \Pi
_{\alpha\sigma} + D_{\sigma\alpha}^{-1}D_{\alpha\sigma}^{-1} D_{\alpha\alpha
}^{2}\Pi_{\alpha\alpha} \right)  G_{\sigma\sigma}\;.
\end{equation}
The second term and the two first terms in the second set of parentheses
already yield the right-hand side of Eq.\ (\ref{identity}). We thus have to
show that the remaining terms cancel up to the order we are computing.

Let us first look at the second term in the first, and the third term in the
second parentheses,
\begin{equation}
\label{Zwischen2}\Pi_{\alpha\sigma}\, \frac{G_{\sigma\alpha}^{-1}}%
{G_{\sigma\sigma}^{-1}} G_{\alpha\alpha} - \frac{D_{\sigma\alpha}^{-1}%
}{D_{\alpha\alpha}^{-1}}\, \Pi_{\alpha\sigma}\, G_{\sigma\sigma} = \Pi
_{\alpha\sigma}\left(  \frac{G_{\sigma\alpha}^{-1}}{G_{\alpha\alpha}^{-1}} -
\frac{D_{\sigma\alpha}^{-1}}{D_{\alpha\alpha}^{-1}} \right)  G_{\sigma\sigma
}\;,
\end{equation}
where we have used Eq.\ (\ref{Gmatrix2}) to replace $G_{\alpha\alpha
}/G_{\sigma\sigma}^{-1}$ by $G_{\sigma\sigma}/G_{\alpha\alpha}^{-1}$. To
two-loop order, we may now safely approximate $G_{\sigma\alpha}^{-1}%
/G_{\alpha\alpha}^{-1}$ by $D_{\sigma\alpha}^{-1}/D_{\alpha\alpha}^{-1}$, and
we see that the expression (\ref{Zwischen2}) vanishes. The remaining terms in
Eq.\ (\ref{Zwischen3}), which we have to consider, are
\begin{equation}
\label{Zwischen4}\left(  D_{\alpha\alpha}-G_{\alpha\alpha} +D_{\sigma\alpha
}^{-1}D_{\alpha\sigma}^{-1} D_{\alpha\alpha}^{2} G_{\sigma\sigma} \right)
\Pi_{\alpha\alpha} \;.
\end{equation}
To two-loop order, we may replace
\begin{equation}
\frac{D_{\sigma\alpha}^{-1}D_{\alpha\sigma}^{-1}}{D_{\alpha\alpha}^{-1}}
\simeq\frac{G_{\sigma\alpha}^{-1}G_{\alpha\sigma}^{-1}}{G_{\alpha\alpha}^{-1}}
\equiv G_{\sigma\sigma}^{-1} - \left[  G_{\sigma\sigma}\right]  ^{-1}\;,
\end{equation}
where we have used the (inverse of the) second Eq.\ (\ref{Gmatrix2}).
Inserting this into Eq.\ (\ref{Zwischen4}), we obtain
\begin{align}
\left(  D_{\alpha\alpha}-G_{\alpha\alpha} +D_{\sigma\alpha}^{-1}%
D_{\alpha\sigma}^{-1} D_{\alpha\alpha}^{2} G_{\sigma\sigma} \right)
\Pi_{\alpha\alpha}  &  \simeq\left(  D_{\alpha\alpha}-G_{\alpha\alpha}
+D_{\alpha\alpha} \left\{  G_{\sigma\sigma}^{-1} - \left[  G_{\sigma\sigma
}\right]  ^{-1} \right\}  G_{\sigma\sigma} \right)  \Pi_{\alpha\alpha
}\nonumber\\
&  = \left(  -G_{\alpha\alpha} + \frac{G_{\sigma\sigma}^{-1}}{D_{\alpha\alpha
}^{-1}}\, G_{\sigma\sigma}\right)  \Pi_{\alpha\alpha} \simeq\left(
-G_{\alpha\alpha} + \frac{G_{\sigma\sigma}^{-1}}{G_{\alpha\alpha}^{-1}}\,
G_{\sigma\sigma}\right)  \Pi_{\alpha\alpha}\;,
\end{align}
where we have again made use of $D_{\alpha\alpha}^{-1} \simeq G_{\alpha\alpha
}^{-1}$ (which is correct up to the order we are computing). The right-hand
side of this equation vanishes on account of the first two
Eqs.\ (\ref{Gmatrix2}). We have thus proved the validity of
Eq.\ (\ref{identity}) up to two-loop order.

All one-loop terms in Eq.\ (\ref{veff2}) can now be written as
\begin{align}
&  \frac{1}{2}\int_{K}\left[  \ln G_{\alpha\alpha}^{-1}(K)+\ln[G_{\sigma
\sigma}(K)]^{-1}+(N-1)\ln G_{\pi\pi}^{-1}(K)\right. \nonumber\\
&  +\left.  D_{\alpha\alpha}^{-1}G_{\alpha\alpha}(K)+D_{\alpha\sigma}%
^{-1}(\phi)G_{\sigma\alpha}(K)+D_{\sigma\alpha}^{-1}(\phi)G_{\alpha\sigma
}(K)+D_{\sigma\sigma}^{-1}(K;\alpha_{0})G_{\sigma\sigma}(K)+(N-1)D_{\pi\pi
}^{-1}(K;\alpha_{0})G_{\pi\pi}(K)-(N+1)\right] \nonumber\\
&  \simeq\frac{1}{2}\int_{K}\left\{  \ln[G_{\sigma\sigma}(K)]^{-1}+(N-1)\ln
G_{\pi\pi}^{-1}(K)\right. \nonumber\\
&  \hspace*{1cm}+\left.  \left[  D_{\sigma\sigma}^{-1}(K;\alpha_{0}%
)-\frac{D_{\sigma\alpha}^{-1}(\phi)\,D_{\alpha\sigma}^{-1}(\phi)}%
{D_{\alpha\alpha}^{-1}}\right]  \,G_{\sigma\sigma}(K)+(N-1)D_{\pi\pi}%
^{-1}(K;\alpha_{0})G_{\pi\pi}(K)-N\right\} \nonumber\\
&  =\frac{1}{2}\int_{K}\left\{  \ln[G_{\sigma\sigma}(K)]^{-1}+(N-1)\ln
G_{\pi\pi}^{-1}(K)\right. \nonumber\\
&  \hspace*{1cm}+\left.  \left[  -K^{2}+\dfrac{2}{N\varepsilon}\left(
3\phi^{2}-\upsilon_{0}^{2}\right)  \right]  G_{\sigma\sigma}(K)+(N-1)\left[
-K^{2}+\dfrac{2}{N\varepsilon}\left(  \phi^{2}-\upsilon_{0}^{2}\right)
\right]  G_{\pi\pi}(K)-N\right\} \nonumber\\
&  +\dfrac{2}{N\varepsilon}\left\{  \int_{K}\left[  G_{\sigma\sigma
}(K)+(N-1)G_{\pi\pi}(K)\right]  \right\}  ^{2}\;. \label{oneloopi}%
\end{align}
Finally, we consider $V_{2}(G)$, cf.\ Eq.\ (\ref{V2i}), for the stationary
values of the two-point functions involving the $\alpha$ field. To two-loop
order, it is sufficient to replace all these functions by the corresponding
expressions given in Eq.\ (\ref{Gmatrix1loopcasei}), resulting in
\begin{align}
V_{2}(G)  &  \simeq\frac{1}{N\varepsilon}\left[  \int_{K}G_{\sigma\sigma
}(K)\right]  ^{2}+\frac{N-1}{N\varepsilon}\left[  \int_{K}G_{\pi\pi
}(K)\right]  ^{2}\nonumber\\
&  -\left(  \dfrac{2\phi}{N\varepsilon}\right)  ^{2}\int_{K}\int_{P}G_{\sigma
}(K+P)\left[  3\,G_{\sigma}(K)G_{\sigma}(P)+(N-1)G_{\pi}(K)G_{\pi}(P)\right]
\;. \label{2loopi}%
\end{align}
Adding Eqs.\ (\ref{treelevel}), (\ref{oneloopi}), and (\ref{2loopi}), we see
that the effective potential becomes identical to the one in the standard
linear $\sigma$ model, Eq.\ (\ref{v2cjt}).

\section{Renormalization}
\label{renorm}

In this appendix, we demonstrate how to renormalize our linear $\sigma$ model
within the auxiliary-field method in one-loop approximation. There is a rich
literature on this subject: the renormalization of scalar field theories
within $\Phi-$derivable approximation schemes was, to our knowledge for the
first time, demonstrated in Ref.\ \cite{Blaizot:2003an}. An iterative
renormalization scheme for the $O(N)$ model in the $1/N$ expansion was
developed in Ref.\ \cite{Berges:2005hc}. This was scheme was applied to pion
and kaon condensation in Ref.\ \cite{Andersen:2006ys}. In
Refs.\ \cite{Cooper:2005vw,Jakovac:2008zq} the $O(N)$ model was renormalized
using the $1/N$ expansion within the auxiliary-field method. Here, we follow
Ref.\ \cite{Fejos:2007ec} where a one-step approach to renormalization of
$\Phi-$derivable approximations was introduced and shown to be equivalent to
the iterative renormalization scheme of the above works. We mention that this
one-step approach was also used in Ref.\ \cite{Fejos:2009dm} for the
renormalization of the $O(N)$ model using the $1/N$ expansion both with and
without the auxiliary-field method.

In order to renormalize Eqs.\ (\ref{con}), (\ref{Msigma}), and (\ref{Mpi}) to
one-loop order it is sufficient to add the following 
five counter terms to the tree-level
potential $U(\phi,\alpha_{0})$:
\begin{equation}
\dfrac{1}{2}\,\delta Z_{1}\,i\alpha_{0}\,\phi^{2}-\dfrac{1}{2}\,\delta
Z_{2}\,i\alpha_{0}\,\upsilon_{0}^{2}+\frac{N\varepsilon}{8}\,\delta
Z_{3}\,\alpha^{2}+\frac{\delta Z_{4}}{2}\,\phi^{2}+\frac{\delta Z_{5}}%
{4}\,\phi^{4}\text{ ,}%
\end{equation}
such that
\begin{equation}
U(\phi,\alpha_{0})\longrightarrow U_{CT}(\phi,\alpha_{0})=\dfrac{i}{2}%
\,Z_{1}\,\alpha_{0}\,\phi^{2}-\dfrac{i}{2}\,Z_{2}\,\alpha_{0}\,\upsilon
_{0}^{2}+\frac{N\varepsilon}{8}\,Z_{3}\,\alpha^{2}+\frac{\delta Z_{4}}%
{2}\,\phi^{2}+\frac{\delta Z_{5}}{4}\,\phi^{4}\text{ ,}%
\end{equation}
where $Z_{i}=1+\delta Z_{i}\,\;i=1,2,3$. Equations (\ref{con}) and (\ref{Mpi})
then read
\begin{align}
h  &  =\phi\left[  Z_{1}\,i\alpha_{0}+\delta Z_{4}+\delta Z_{5}\phi^{2}%
+\dfrac{4}{N\varepsilon}\int_{K}G_{\sigma}(K)\right]  \text{ }, \label{conren}%
\\
M_{\pi}^{2}  &  =i\alpha_{0}=\dfrac{2}{Z_{3}\,N\varepsilon}\left[  Z_{1}%
\,\phi^{2}-Z_{2}\,\upsilon_{0}^{2}+\int_{K}G_{\sigma}(K)+(N-1)\int_{K}G_{\pi
}(K)\right]  \text{ . } \label{Mpiren}%
\end{align}
Using a cut-off $\Lambda_{\rm CO}$ for the four-dimensional momentum
integration (and neglecting terms of order 
$\mu^2/\Lambda_{\rm CO}^2$, where $\mu$ is the renormalization scale) 
the tadpole integrals can be written as
\cite{Fejos:2007ec}
\begin{equation}
\int_{K}G_{i}(K)=\Lambda^{2}+T_{d}M_{i}^{2}+T_{F}^{i}\text{ ,} \label{tadpole}%
\end{equation}
where $\Lambda= 4 \pi \Lambda_{\rm CO}$,
\[
T_{d}=-\frac{1}{16\pi^{2}}\, \ln   \frac{16\pi^{2}\Lambda^{2}}{\mu^{2}e}\;,
\]
and
\begin{equation}
T_{F}^{i}=\int\frac{d^{3}\vec{k}}{\left(  2\pi\right)  ^{3}}\;\frac{1}%
{\sqrt{\vec{k}^{\;2}+M_{i}^{2}}}\left[  \exp\left(  \sqrt{\vec{k}^{\;2}%
+M_{i}^{2}}/T\right)  -1\right]  ^{-1}+\frac{1}{16\pi^{2}}\left(
M_i^2\ln \frac{M_{i}^{2}}{\mu^{2}} -M_i^2 + \mu^2 \right)\;,\;\;i=\sigma,\pi\;. \label{Tf}%
\end{equation}
Inserting Eq.\ (\ref{tadpole}) into Eq.\ (\ref{Mpiren}), we obtain
\begin{align}
M_{\pi}^{2}  &  =\frac{2}{N\varepsilon}\left(  \dfrac{Z_{1}}{Z_{3}}\,\phi
^{2}-\dfrac{Z_{2}}{Z_{3}}\,\upsilon_{0}^{2}+\dfrac{1}{Z_{3}}\left\{
N\Lambda^{2}+T_{F}^{\sigma}+\left(  N-1\right)  T_{F}^{\pi}+\frac{2T_{d}%
}{\varepsilon}\left[  \frac{N+2}{N}\phi^{2}-\upsilon_{0}^{2}+T_{F}^{\sigma
}+\left(  N-1\right)  T_{F}^{\pi}\right]  \right\}  \right) \nonumber\\
&  =\frac{2}{N\varepsilon}\left[  \phi^{2}-\upsilon_{0}^{2}+T_{F}^{\sigma
}+(N-1)T_{F}^{\pi}\right] \nonumber\\
&  +\frac{2}{N\varepsilon}\left\{  \left(  \dfrac{Z_{1}}{Z_{3}}-1\right)
\,\phi^{2}-\left(  \dfrac{Z_{2}}{Z_{3}}-1\right)  \,\upsilon_{0}^{2}+\dfrac
{1}{Z_{3}}\left\{  N\Lambda^{2}+\frac{2T_{d}}{\varepsilon}\left[  \frac
{N+2}{N}\phi^{2}-\upsilon_{0}^{2}+T_{F}^{\sigma}+\left(  N-1\right)
T_{F}^{\pi}\right]  \right\}  \right. \nonumber\\
&  \hspace*{1cm}+\left.  \left(  \dfrac{1}{Z_{3}}-1\right)  \left[
T_{F}^{\sigma}+\left(  N-1\right)  T_{F}^{\pi}\right]  \right\}  \;.
\end{align}
The first line is the expected, finite result for the pion mass (squared). The
renormalization constants have to be chosen such that the second and third
lines vanish. Cancellation of the temperature-dependent sub-divergence [the
terms proportional to $T_{F}^{\sigma}+(N-1)T_{F}^{\pi}$] requires
\begin{equation}
Z_{3}=1+\frac{2T_{d}}{\varepsilon}\;\;\Longleftrightarrow\;\;\delta
Z_{3}=\frac{2T_{d}}{\varepsilon}\;. \label{Z_3}%
\end{equation}
Cancellation of the $\phi-$dependent overall divergence gives
\begin{equation}
\frac{Z_{1}}{Z_{3}}-1=-\frac{2T_{d}}{\varepsilon\,Z_{3}}\,\frac{N+2}%
{N}\;\;\Longleftrightarrow\;\;Z_{1}=1-\frac{4T_{d}}{N\varepsilon
}\;\;\;\;\Longleftrightarrow\;\;\delta Z_{1}=-\frac{4T_{d}}{N\varepsilon}\;,
\label{Z_1}%
\end{equation}
where we have used the result (\ref{Z_3}) for $Z_{3}$. Finally, cancellation
of the constant overall divergence yields
\begin{equation}
\frac{N\Lambda^{2}}{Z_{3}}=\upsilon_{0}^{2}\left(  \frac{Z_{2}}{Z_{3}}%
-1+\frac{2T_{d}}{\varepsilon\,Z_{3}}\right)  \;\;\Longleftrightarrow
\;\;Z_{2}=1+\frac{N\Lambda^{2}}{\upsilon_{0}^{2}}\;\;\Longleftrightarrow
\;\;\delta Z_{2}=\frac{N\Lambda^{2}}{\upsilon_{0}^{2}}\;,
\end{equation}
where we have again used Eq.\ (\ref{Z_3}). Finally, turning to
Eq.\ (\ref{conren}), we can use $M_{\pi}^{2}=i\alpha_{0}$ and
Eq.\ (\ref{tadpole}) to write
\begin{equation}
h=\phi\left[  Z_{1}\,M_{\pi}^{2}+\delta Z_{4}+\delta Z_{5}\,\phi^{2}+\frac
{4}{N\varepsilon}\left(  \Lambda^{2}+T_{d}\,M_{\sigma}^{2}+T_{F}^{\sigma
}\right)  \right]  \;.
\end{equation}
Using the result (\ref{Z_1}) for $Z_{1}$, we obtain
\begin{equation}
h=\phi\left[  M_{\pi}^{2}+\frac{4}{N\varepsilon}\,T_{F}^{\sigma}+\delta
Z_{4}+\frac{4\Lambda^{2}}{N\varepsilon}+\delta Z_{5}\phi^{2}+\frac{4T_{d}%
}{N\varepsilon}\left(  M_{\sigma}^{2}-M_{\pi}^{2}\right)  \right]  \;.
\end{equation}
The first two terms represent the expected, finite result. The counter terms
$\delta Z_{4,5}$ have to be chosen such that the remaining (infinite) terms
cancel. Using the fact that
\begin{equation}
M_{\sigma}^{2}=M_{\pi}^{2}+\frac{4\phi^{2}}{N\varepsilon}\;,
\end{equation}
we see that this is achieved by the choice
\begin{align}
\delta Z_{4}  &  =-\frac{4\Lambda^{2}}{N\varepsilon}\;,\\
\delta Z_{5}  &  =-\frac{16T_{d}}{N^{2}\varepsilon^{2}}\;.
\end{align}
This completes the renormalization of the linear $\sigma$ model in one-loop
approximation within the auxiliary-field method.

Thus, to one-loop order the renormalized equations for the condensate and for
the masses read%
\begin{align}
h &  =\frac{2\phi}{N\varepsilon}\left[  \phi^{2}-\upsilon_{0}^{2}+3T_{F}^{\sigma
}+(N-1)T_{F}^{\pi}\right]  \;,\label{phiren}\\
M_{\pi}^{2} &  =\frac{2}{N\varepsilon}\left[  \phi^{2}-\upsilon_{0}^{2}%
+T_{F}^{\sigma}+(N-1)T_{F}^{\pi}\right]  \;,\label{mpiren}\\
M_{\sigma}^{2} &  =\frac{2}{N\varepsilon}\left[  3\phi^{2}-\upsilon_{0}%
^{2}+T_{F}^{\sigma}+(N-1)T_{F}^{\pi}\right]  \text{ .}\label{msiren}%
\end{align}
where $T_{F}^{i}$ is given by Eq.\ (\ref{Tf}).%

\addcontentsline{toc}{section}{Abbildungs- und Literaturverzeichnis}%

\end{document}